\documentclass[floatfix, twocolumn,showpacs,nofootinbib]{revtex4}
\usepackage{epsf}
\usepackage{graphicx}% Include figure files
\usepackage{dcolumn}% Align table columns on decimal point

\newcommand{\BE}{\begin{equation}}
\newcommand{\EE}{\end{equation}}
\newcommand{\BA}{\begin{eqnarray}}
\newcommand{\EA}{\end{eqnarray}}

\newcommand{\etal}{{\it et al.}}

\def\be{\begin{equation}}
\def\ee{\end{equation}}
\def\bea{\begin{eqnarray}}
\def\eea{\end{eqnarray}}

\def\fun#1#2{\lower3.6pt\vbox{\baselineskip0pt\lineskip.9pt
        \ialign{$\mathsurround=0pt#1\hfill##\hfil$\crcr#2\crcr\sim\crcr}}}

\begin{document}
\input epsf
\renewcommand{\topfraction}{0.8}
\preprint{astro-ph/yymmnnn, \today}

\title{Probing Cosmology with Weak Lensing Minkowski Functionals}
\author{Jan M. Kratochvil$^{1}$, Eugene A. Lim$^{2,6}$, Sheng Wang$^{3}$, Zolt\'an Haiman$^{2,4}$, Morgan May$^{5}$, Kevin Huffenberger$^{1}$}
\affiliation{{$^1$ Department of Physics, University of Miami, 1320 Campo Sano Drive, Coral Gables, FL 33143, USA}}
\affiliation{{$^2$ Institute for Strings, Cosmology, and Astroparticle Physics (ISCAP), Columbia University, 550 West 120th Street, New York, NY 10027, USA}}  
\affiliation{{$^3$ Kavli Institute for Cosmological Physics, University of Chicago, 933 East 56th Street, Chicago, IL 60637, USA}}
\affiliation{{$^4$ Department of Astronomy and Astrophysics, Columbia University, 550 West 120th Street, New York, NY 10027, USA}} 
\affiliation{{$^5$ Physics Department, Brookhaven National Laboratory, Upton, NY 11973, USA}   }
\affiliation{{$^6$ Dept. of Applied Mathematics and Theoretical Physics, University of Cambridge, Wilberforce Road, CB3 0WA Cambridgeshire, UK}}
\begin{abstract}
  In this paper, we show that Minkowski Functionals (MFs) of weak
  gravitational lensing (WL) convergence maps contain significant non-Gaussian,
  cosmology-dependent information. To do this, we use a large suite of
  cosmological ray-tracing N-body simulations to create mock weak WL convergence maps,
  and study the cosmological information content of MFs derived from these maps.
  Our suite consists of 80 independent $512^3$ N-body runs, covering
  seven different cosmologies, varying three cosmological parameters
  $\Omega_m$, $w$, and $\sigma_8$ one at a time, around a fiducial
  $\Lambda$CDM model.  In each cosmology, we use ray-tracing to create
  a thousand pseudo-independent 12 ${\rm deg}^2$ convergence maps, and
  use these in a Monte Carlo procedure to estimate the joint
  confidence contours on the above three parameters.  We include
  redshift tomography at three different source redshifts $z_s=1, 1.5,
  2$, explore five different smoothing scales $\theta_G=1, 2, 3, 5,
  10$ arcmin, and explicitly compare and combine the MFs with the WL
  power spectrum.
  We find that the MFs capture a substantial amount of information
  from non-Gaussian features of convergence maps, i.e. beyond the
  power spectrum. The MFs are particularly well suited to break
  degeneracies and to constrain the dark energy equation of state
  parameter $w$ (by a factor of $\approx$ three better than from the
  power spectrum alone). The non-Gaussian information derives partly
  from the one-point function of the convergence (through $V_0$, the
  ``area'' MF), and partly through non-linear spatial information
  (through combining different smoothing scales for $V_0$, and through
  $V_1$ and $V_2$, the boundary length and genus MFs, respectively).
  In contrast to the power spectrum, the best constraints from the
  MFs are obtained only when multiple smoothing scales are combined.

\end{abstract}
% \vskip2pc]
\maketitle

%%%%%%%%%%%%%%%%%%%%%%%%%%%%%%%%%%%%%%%%%%%%%%%%%%%%%%%%%%%%%%%%%%%%%%%%%%%%%%%%%%%%%%%%%%
\section{Introduction}
%%%%%%%%%%%%%%%%%%%%%%%%%%%%%%%%%%%%%%%%%%%%%%%%%%%%%%%%%%%%%%%%%%%%%%%%%%%%%%%%%%%%%%%%%%

Forthcoming large weak gravitational lensing (WL) surveys, such as
those by the \textit{Large Synoptic Survey Telescope} (LSST), will
provide a growing number of cosmic shear measurements of increasing quality.
This prospect poses the theoretical challenges of (i) how much
statistical information could be extracted from the expected WL maps,
and (ii) to what extent this information will help constrain
cosmological models.  The power spectrum, or equivalently the
two-point correlation function, are of unquestionable importance, but
are nevertheless incomplete statistical descriptions of the lensing
maps.  This is because gravity is non-linear, and it generates
non-Gaussianity on small scales.  Perhaps the most natural way to
proceed and characterize non-Gaussian statistical signatures is by
using higher--order poly-spectra \cite{Zaldarriaga:2002qt}, or
equivalently, three--point and higher--order correlation functions
\cite{Bernardeau97,Hui99, T&J02, Takada:2003sv,T&J04, Sefusatti06,
Berge:2009xj,Takada:2008fn, Jain:2003tba}.  An interesting, and less
explored alternative, originally suggested in the context of the
cosmic density field~\cite{gmd86}, is to utilize topological features.

As an example of this approach, the genus of different iso-density
contour surfaces~\cite{hgw86,gwm87,g89,v94,pkg05} has recently been
measured in the \textit{Sloan Digital Sky Survey} (SDSS).  The genus
curve---i.e.\ genus as a function of density threshold---was derived
for both the luminous red galaxy (LRGs) and the main galaxy samples,
and found to deviate from the theoretical prediction for a Gaussian
random field.  These deviations then allowed a testing of different
galaxy formation scenarios, through their nonlinear biasing and other
gravitational effects~\cite{g09,c10}.

The genus is only one of several topological measures of iso-density
surfaces (in three-dimensions) or contours (in two-dimensions), which
are collectively known as the Minkowski functionals
(MFs)~\cite{mbw94}.  The full set of MFs has been successfully applied
to maps of Cosmic Microwave Background (CMB) temperature anisotropies,
to probe primordial non-Gaussianity.  There one compares the MFs
derived from the data to those expected in standard inflationary
models with Gaussian initial conditions ~\cite{WMAP1,WMAP3,WMAP5}.
The current constraints from the MFs on the commonly used
non-Gaussianity parameter $f_{\rm NL}$ are comparable to those from
the bispectrum \cite{Hikage:2008gy}.  However, the MFs have the
additional ability to test models of the early universe with
non-Gaussianity only appearing at the level of trispectrum or
beyond~\cite{m10}.

Motivated by these demonstrations that MFs are powerful yet simple
probes of non-Gaussianity, we here perform a detailed and systematic
study the MFs of the cosmic shear field and their potential use to
differentiate between cosmological models.  Related previous works
have considered the $\Omega_m$--dependence of the one--point function
and WL peak statistics, using ray-tracing
simulations~\cite{JSW00,BJ&VW00}, and have demonstrated the ability of
WL MFs to distinguish standard cold dark matter (SCDM), open CDM and
$\Lambda$CDM models~\cite{Sato:2001cb,g02}. It has also been shown
that one of the MFs, the area statistic, could place strong
constraints on the properties of dark energy~\cite{whm09}.  Finally, a
recent theoretical analysis presents a perturbative description of WL
MFs \cite{Munshi:2011wu}.

Our work here extends the previous studies in many ways, including the
scope, detail, and the level of interpretation of the results.  In
particular, we have run a large suite of $N$-body simulations with
ray-tracing to generate WL maps for seven different cosmological
models.  We vary three cosmological parameters, $\Omega_m$, $w$, and
$\sigma_8$, one at a time, around a fiducial model.  Using our
simulation suite, we are able to compute the joint confidence levels
on these three parameters through a Monte Carlo procedure.  We
quantify in detail where these constraints come from, finding ideal
combinations of galaxy source redshifts and smoothing scales. Finally,
in order to isolate the non-Gaussian information, we explicitly
compare and combine the constraints from the MFs with those from the
WL power spectrum.

The core finding of this paper is that the MFs capture significant
non-Gaussian, cosmology-dependent information.  In particular, the
dark energy equation of state parameter $w$ is constrained by a factor
of approximately three better than from the power spectrum alone. We
attribute this improvement to the MFs breaking degeneracies between
$w$ and the other parameters.  We also explore combinations of the
three different MFs, and of the MFs with the power spectrum. This
allows us to assess the amount of additional information contained in
the MFs (beyond the power spectrum), and to interpret the nature of
the non-Gaussian information.

The rest of this paper is organized as follows.  The details of our
simulations and mock WL maps are described in
Section~\ref{Simulations}.  The basics of MFs and the algorithm to
measure them from our simulated maps are described in
Section~\ref{Minkowski Functionals}, while Section~\ref{Power Spectra}
describes our use of the WL power spectrum. Our Monte Carlo procedure,
using the maps to compute the confidence levels on the parameters, is
described in Section~\ref{Statistics}, along with a discussion of the
required number of pseudo-independent realizations.
Section~\ref{Results} presents our main results, showing the
constraints on the cosmological parameters.  Various combinations of
the observables are studied: using individual MFs separately,
combining MFs with the power spectrum, using redshift tomography,
combining smoothing scales, using maps with and without intrinsic
ellipticity noise, etc. In this section, we also compare our
power-spectrum-alone constraints to the literature, and discuss
several possible sources of inaccuracy, some quantified here, as well
as some left for future work. Finally, in Section~\ref{Conclusions},
we summarize our conclusions and the implications of this work.

%%%%%%%%%%%%%%%%%%%%%%%%%%%%%%%%%%%%%%%%%%%%%%%%%%%%%%%%%%%%%%%%%%%%%%%%%%%%%%%%%%%%%%%%%%
\section{Simulation Suite and Weak Lensing Maps}\label{Simulations}
%%%%%%%%%%%%%%%%%%%%%%%%%%%%%%%%%%%%%%%%%%%%%%%%%%%%%%%%%%%%%%%%%%%%%%%%%%%%%%%%%%%%%%%%%%

\subsection{N-body Simulations}

The large-scale structure simulations and lensing maps were created
with our new Inspector Gadget lensing simulation pipeline
\cite{InspectorGadget1, InspectorGadget2} on the New York Blue
supercomputer, which is part of the New York Center for Computational
Sciences at Brookhaven National Laboratory/Stony Brook University. The
center features an IBM Blue Gene/L with 36,864 CPUs and a Blue Gene/P
with 8,192 CPUs. We ran a series of 80 CDM $N$-body simulations with
$512^3$ particles each and a box size of $240h^{-1}$~Mpc. This
corresponds to a three-week test run for our pipeline. We used a
modified version of the public N-body code Gadget-2
\cite{Springel:2005mi}, which we adapted for the Blue Gene/L and /P,
and enhanced to allow the dark energy equation of state parameter
$w\neq 1$, as well as to compute weak lensing related quantities, such
as comoving distances to the observer.  Gadget's adjustable parameters
were fine-tuned for maximum throughput on the cluster. Volker Springel
kindly provided us with the initial conditions (IC) generator
N-GenIC. The total linear matter power spectrum, which is an input for
the IC generator, was created with the Einstein-Boltzmann code CAMB
\cite{Lewis:1999bs} for $z=0$, and scaled to the starting redshift of
our simulations at $z=100$ following the linear growth factor. The WL
maps were created, and the rest of the analysis were performed with
our own proprietary codes and file formats. Up to 70TB of simulation
and lensing products were stored in the process of this work.

A total of 80 N-body simulations were run, covering 7 different
cosmological models in multiple runs with different random initial
conditions. A total of 50 of the runs were performed in our fiducial
cosmology, with parameters chosen to be \{$\Omega_m=0.26$,
$\Omega_\Lambda=0.74$, $w=-1.0$, $n_s=0.96$, $\sigma_8=0.798$,
$H_0=0.72\}$.  These runs all used the same input power spectrum, but
a different and strictly independent realization, yielding a
statistically robust set of maps, and allowing us to assess how much
the use of fewer independent realizations (used in the other models)
affects the results.  In each of the other six cosmological models, we
varied one parameter at a time, keeping the others fixed, with the
following values: $\Omega_m=\{0.23, 0.29\}$ (while
$\Omega_\Lambda=\{0.77, 0.71\}$ such that the universe stays spatially
flat), $w=\{-0.8, -1.2\}$, and $\sigma_8=\{0.75, 0.85\}$. For these
six non-fiducial cosmological models we ran 5 simulations each, with a
different realization of the initial conditions.\footnote{We also ran
  ten additional simulations, bracketing the scalar spectral index
  $n_s=\{0.92, 1.00\}$, but were unable to reliably distinguish these
  from the fiducial model.  These runs will therefore not be described
  further in this paper, and the scalar spectral index will be fixed
  at its fiducial value of $n_s=0.96$.}

\begin{table}
\begin{tabular}{|l|c|c|c|c|c|} 
\hline
Identifier & $\sigma_8$ & $w$ & $\Omega_m$ & $\Omega_\Lambda$ & \# of sims \\
\hline
Fiducial & 0.798 & -1.0 & 0.26 & 0.74 &45\\
Auxiliary & 0.798 & -1.0 & 0.26 & 0.74 & 5\\
Om23 & 0.798 & -1.0 & 0.23 & 0.77 & 5\\
Om29 & 0.798 & -1.0 & 0.29 & 0.71& 5\\
w12 & 0.798 & -1.2 & 0.26 & 0.74 & 5\\
w08 & 0.798 & -0.8 & 0.26 & 0.74 & 5\\
si75 & 0.750 & -1.0 & 0.26 & 0.74 & 5\\
si85 & 0.850 & -1.0 & 0.26 & 0.74 & 5\\
\hline
\end{tabular}
\caption[]{\textit{Cosmological parameters varied in each model.}}\label{tab:Cosmologies}
\end{table}

\subsection{Lensing Maps}

The weak lensing maps were created with a standard two-dimensional
ray-tracing algorithm.  We refer the reader to our previous work
\cite{Kratochvil:2009wh} and to \cite {InspectorGadget1} for the full
description of our methodology.  The code developed and used in that
paper was adapted here for the Blue Gene/P and /L and parallelized to
deal with the stringent memory requirements, but in its core equations
it remained the same, so we will list here only the specifications as
well as any changes and additions we have made.

We chose to implement the ray-tracing algorithm described in
\cite{Hamana:2001vz}. Earlier work with similar algorithms include
\cite{Schneider:1992, Wambsgaans:1998, Jain:1999ir}. The large-scale
structure from the N-body simulations was output as particle positions
in boxes at different redshifts, starting at redshift $z=2$.  The
particles were then projected onto planes perpendicularly to the
planes and the central line of sight of the map.  We used the
triangular shaped cloud (TSC) scheme \cite{Hockney-Eastwood} to place
the particles on a grid on these two-dimensional density planes; the
particle surface density was then converted into the gravitational
potential via the Poisson equation. The algorithm then followed light
rays from the observer back in cosmic time. The deflection angle, as
well as the weak lensing convergence and shear were calculated at each
plane for each light ray. These depend on the first and second
derivatives of the potential, respectively. Between the planes, the
light rays travel in straight lines. For simplicity, in this work we
utilized only the convergence maps; shear maps were also created and
will be used in the future.

The maps were created for $2048\times2048$ light rays, with the lens
planes spaced $80h^{-1}$Mpc apart along the line of sight. The lens
planes themselves recorded the gravitational potential on a finer
$4096\times4096$ grid, to avoid losing power on small angular scales,
as pointed out in \cite{Sato:2009}. We have found a similar fall-off,
and therefore increased the resolution on the density planes from our
previous publication \cite{Kratochvil:2009wh}.

We created one thousand 12-square-degree convergence maps for each
cosmology, by mixing simulations with different random initial
conditions, and by randomly rotating and shifting the simulation data
cubes. For the maps in each non-fiducial cosmology, lens planes mixed
from all five independent N-body runs were used. In the fiducial
cosmology, we created two sets of 1,000 maps. One of these sets was
created from the five independent N-body runs with the same five
quasi-identical\footnote{By ``quasi-identical'', we mean that the
random number seeds to create the initial particle distributions from
the power spectra were kept the same across all cosmological models,
but the normalization of the power spectra themselves was adjusted
such as to yield the desired $\sigma_8$ today in every cosmology. This
adjustment is necessary due to the difference in growth factors
between the models.} initial conditions as in the non-fiducial
cosmologies, and will hereafter be referred to as the \lq\lq
auxiliary\rq\rq\ set.  The second was created by mixing lens planes
from the remaining larger ensemble of 45 independent N-body runs, and
will be referred to as the \lq\lq fiducial\rq\rq\ set.

Results from the fiducial and auxiliary map sets will be compared
below to verify that they do not depend on a particular set of maps
and simulations. Having two independent sets will also be utilized in
our Monte Carlo procedure, which involves a $\chi^2$--minimization.
This crucially requires that the maps for which the best-fit model
parameters are found are independent of the set used to compute the
covariance matrix.

For simplicity, we assumed the source galaxies are confined to planes
at a fixed redshift. The convergence maps were generated for three
different source redshifts, $z_s=1, 1.5$, and $2$. After the raw maps
with the lensing signal were generated, ellipticity noise from the
random orientations of the source galaxies was added to the maps pixel
by pixel.  We assumed the noise is represented by a Gaussian random
field (with a top-hat filter of pixel size; or equivalently a random
lattice noise model, e.g \cite{Mantz08}).  We further assumed a
uniform source galaxy surface density of
$n_{gal}=15$~galaxies/arcmin$^2$ on each source plane (neglecting
other effects, such as shot noise from random galaxy positions, or
magnification of the source galaxies) and a redshift dependent
root-mean-square of the noise in one component of the shear
\cite{S&K04}
\BE\label{gamma noise}
\sigma_\gamma(z)=0.15+0.035z.
\EE
The r.m.s.\ noise for a pair of pixels $\vec x$ and $\vec y$ on the
convergence maps was then taken to be
\BE\label{noise correlation} 
\langle \kappa_{\rm noise}(\vec{x})\kappa_{\rm noise}(\vec{y})\rangle=\frac{\sigma_\gamma^2}{n_{gal}A}\delta_{\vec{x}\vec{y}},
\EE
where $\delta_{\vec{x}\vec{y}}$ is the Kronecker delta function, and
$A$ is the solid angle of a pixel. The galaxy densities adopted for a
single source redshift is fairly low.  This lets us combine the three
redshifts to employ tomography with a total
$n_{gal}=45$~galaxies/arcmin$^2$, a typical value expected for galaxy
surveys with the depth of LSST (e.g.\ \cite{LSST:2008} and Eq.\ (3.7)
in \cite{LSST:2009pq}).  Once noise in each pixel is added, we
smooth the maps with a finite version of a 2D Gaussian filter, with
the kernel around every pixel $\phi_0$,
\BE\label{Gauss-kernel}
W_G(\phi,\phi_0)=\frac{1}{\pi\theta^2_G}\exp\left(-\frac{(\phi-\phi_0)^2}{\theta_G^2}\right),
\EE
truncated at $6\theta_G$.  We employ five different smoothing scales,
$\theta_G = 1, 2, 3, 5, 10$ arcmin.  The smallest smoothing scale
retains the most information but also the most noise.  Hence, the
strategy is to combine MFs with several different smoothing scales to
extract additional information, despite strong correlations between
maps with different $\theta_G$.

In Section~\ref{Power Spectra} below, we will discuss the accuracy of
our simulations pipeline and the convergence power spectrum derived
from the maps.

%%%%%%%%%%%%%%%%%%%%%%%%%%%%%%%%%%%%%%%%%%%%%%%%%%%%%%%%%%%%%%%%%%%%%%%%%%%%%%%%%%%%%%%%%%
\section{Minkowski Functionals and Power Spectra}
\label{Minkowski Functionals and Power Spectra}
%%%%%%%%%%%%%%%%%%%%%%%%%%%%%%%%%%%%%%%%%%%%%%%%%%%%%%%%%%%%%%%%%%%%%%%%%%%%%%%%%%%%%%%%%%
%
\subsection{Minkowski Functionals}
\label{Minkowski Functionals}

Minkowski Functionals are \emph{morphological} statistics of
thresholded smoothed random fields, complementary to the more familiar
hierarchy of correlation functions. If the fields are strictly
Gaussian, then there exist one--to--one mappings between the power
spectrum and the MFs. In the weakly non-Gaussian case, one can also
find an approximate map between the two statistics
\cite{Hikage:2006fe}, or expand the MFs perturbatively as a function
of the power spectrum \cite{Matsubara:2010te,Munshi:2011wu}.  In the
general case, however, MFs encode information from correlation
functions of arbitrarily high order, which is what makes them useful
probes of non-Gaussianities. Since weak lensing convergence maps are
expected to contain small scale non-Gaussian information, they are
particularly well suited to this application.

As mentioned in the Introduction, a few previous attempts have been
made to use morphological statistics to analyze weak lensing maps.
The early works \cite{JSW00} and \cite{BJ&VW00} considered the
$\Omega_m$--dependence of the one--point function and peak statistics,
using ray-tracing simulations, while~\cite{Sato:2001cb} and
~\cite{g02} used MFs to discern between SCDM, OCDM and $\Lambda$CDM
models.  More recently, \cite{whm09} used the fractional area of ``hot
spots'' of a thresholded map as a statistic, while
\cite{Kratochvil:2009wh, Dietrich:2009jq} used counts of peaks
(defined as local maxima) in convergence and shear maps.~\footnote{The
peaks containing most of the cosmological information were found to
have relatively low amplitudes, and are not typically produced by
single collapsed objects \cite{Yang:2011zz}.}  Finally,
\cite{Maturi:2009as} constructed an analytical approximation to the
peak number counts -- their approximate statistic turns out to be the
\emph{genus}, which, as we will see below, is identical to one of the
three MFs.

In general, for a given $D$-dimensional smoothed field, one can
construct $D+1$ Minkowski Functionals $V_j$.  Since we analyze 2D weak
lensing maps in this paper, we restrict ourselves to reviewing MFs in
two dimensions, and refer the interested reader to
\cite{a81,bbks86,mbw94} for a more comprehensive discussion.  The
three MFs in 2D, $V_0$, $V_1$, and $V_2$, measure the area, boundary,
and Euler characteristic, respectively, of the excursion set $A_\nu$
of an image, defined to include the part above a certain threshold
$\nu$.

The convergence is a smooth scalar field $\kappa({\bf x})$.  For a
given threshold $\nu$, the excursion set $A_{\nu}$ is defined as the
set of points ${\bf x}$ with $\kappa({\bf x}) > \nu$.  The area
statistic, $V_0(\nu)$, is the fractional area above the threshold,
\begin{equation}
V_0(\nu) = \int \Theta(\kappa-\nu)~da,
\end{equation}
where $\Theta(\kappa-\nu)$ is the Heaviside step function. The
boundary length statistic, $V_1(\nu)$, is the total length of
iso-density contours. For computational simplicity, we convert it into
an area integral by inserting a delta function and the appropriate
Jacobian,
\begin{equation}
V_1(\nu) = \frac{1}{4}\int |\nabla \kappa|\delta(\kappa-\nu)~da,
\end{equation}
where $\nabla \kappa$ denotes partial derivatives. Finally, the genus
statistic, $V_2(\nu)$, is the integration of the principal curvature
$K$ along the iso-density contours, which we can similarly convert
into the area integral
\begin{equation}
V_1(\nu) = \frac{1}{2\pi}\int |\nabla \kappa|\delta(\kappa-\nu)K~da,
\end{equation}
with
\begin{equation}
K =|\nabla_{\dot{\gamma}} \dot{\gamma}|
\end{equation}
where $\dot{\gamma}$ is the tangent vector along the curve $\gamma$
defining the contour, and $\nabla_{\dot{\gamma}}$ is the covariant
derivative along the curve.

The reason we have re-expressed the MFs as integrals of invariants is
that the integrands reduce to depend solely on the threshold $\nu$ and
the 1st and 2nd order derivatives of the field $\kappa$,
\begin{equation}
V_0(\nu) = \int\Theta(\kappa({\bf x}) -\nu)~ da \label{eqn:V0},
\end{equation}
\begin{equation}
V_1(\nu) = \int\delta(\kappa({\bf x}) -\nu)\sqrt{ \kappa_x^2 + \kappa_y^2}~ da,
\label{eqn:V1}
\end{equation}
and
\begin{equation}
V_2(\nu) = \int\delta(\kappa({\bf x}) -\nu)\frac{2\kappa_x\kappa_y\kappa_{xy}-\kappa_x^2\kappa_{yy}-\kappa_y^2\kappa_{xx}}{ \kappa_x^2 + \kappa_y^2}~ da,
\label{eqn:V2}
\end{equation}
where $\kappa_x$ means partial derivative with respect to $x$, etc.

In this form, numerical calculation of the MFs $V_j(\nu)$ from a
pixelized map becomes simple: we calculate the derivatives in
coordinate space via finite difference and then sum them over the
entire space with its corresponding threshold value $\nu$.

\begin{figure*}[htp]
\centering
\includegraphics[width=8 cm]{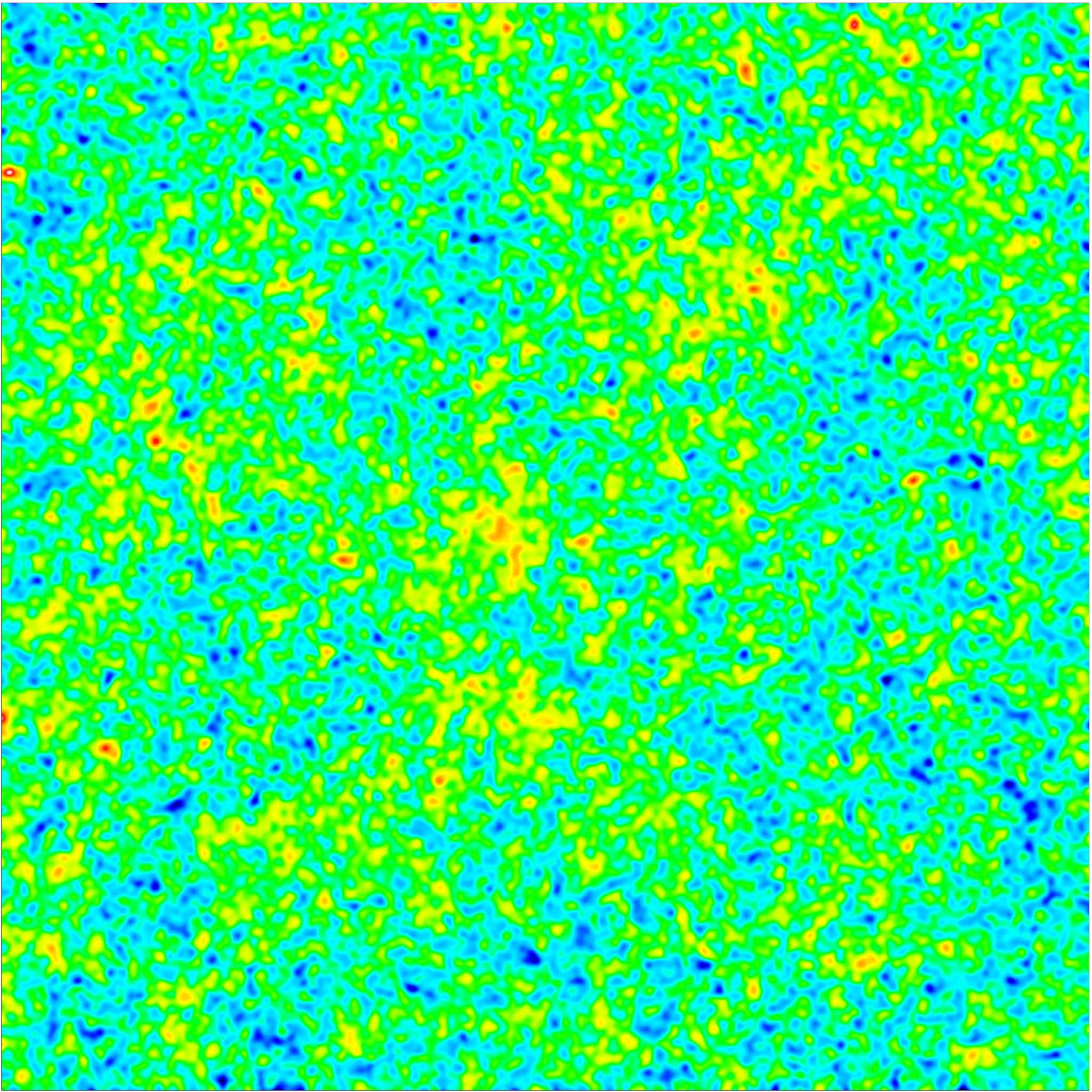} 
\includegraphics[width=8 cm]{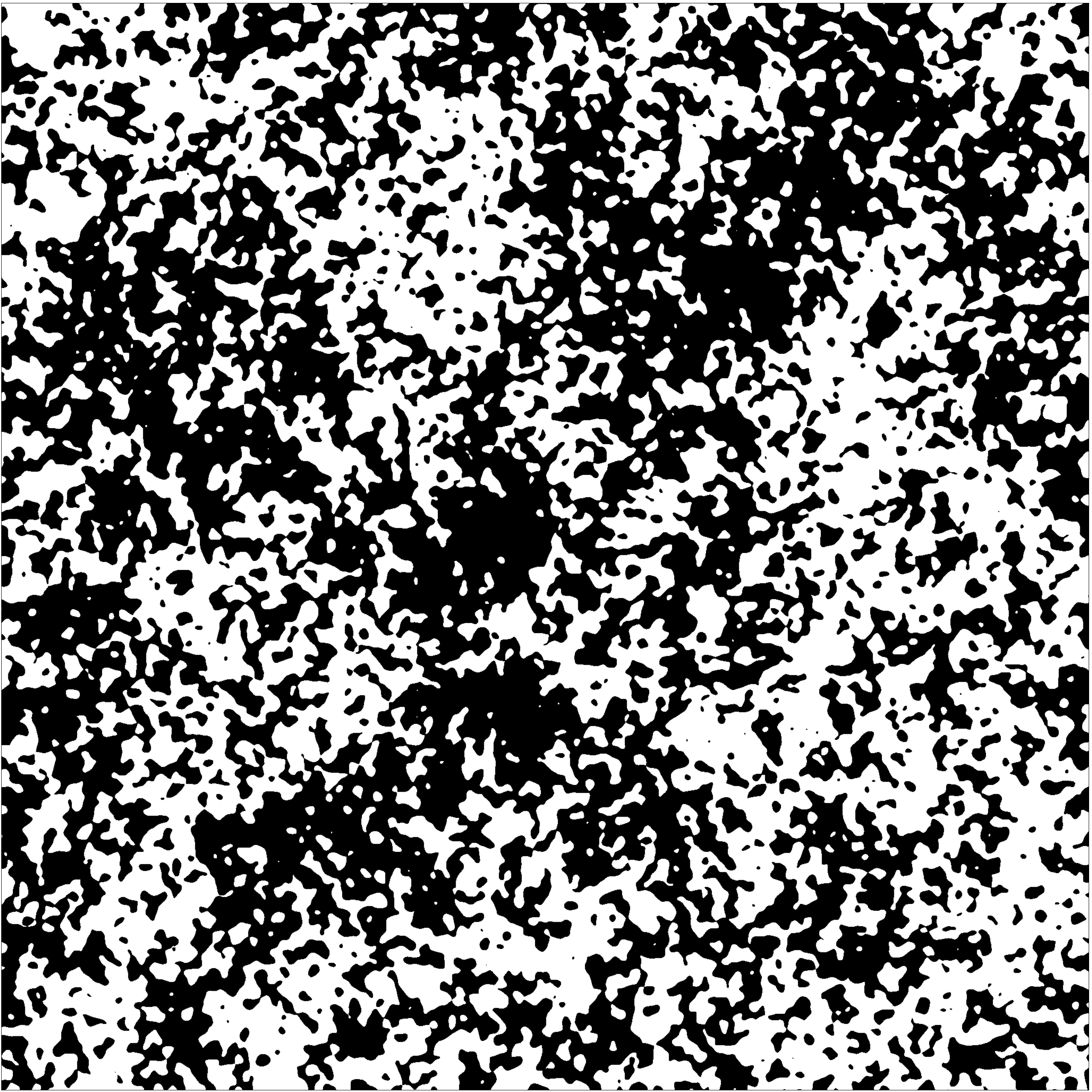}\\
\vspace{0.07cm}
\includegraphics[width=8 cm]{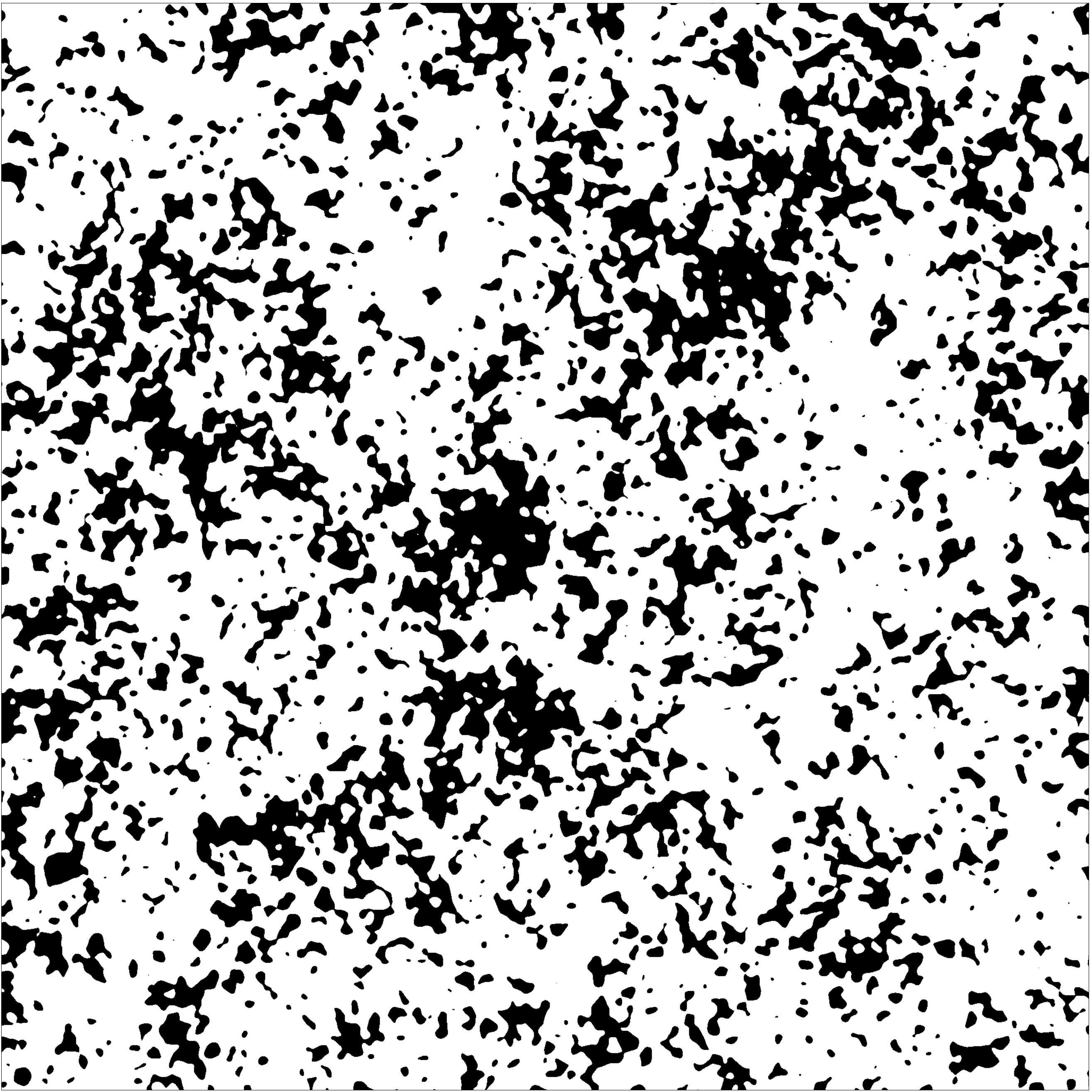} 
\includegraphics[width=8 cm]{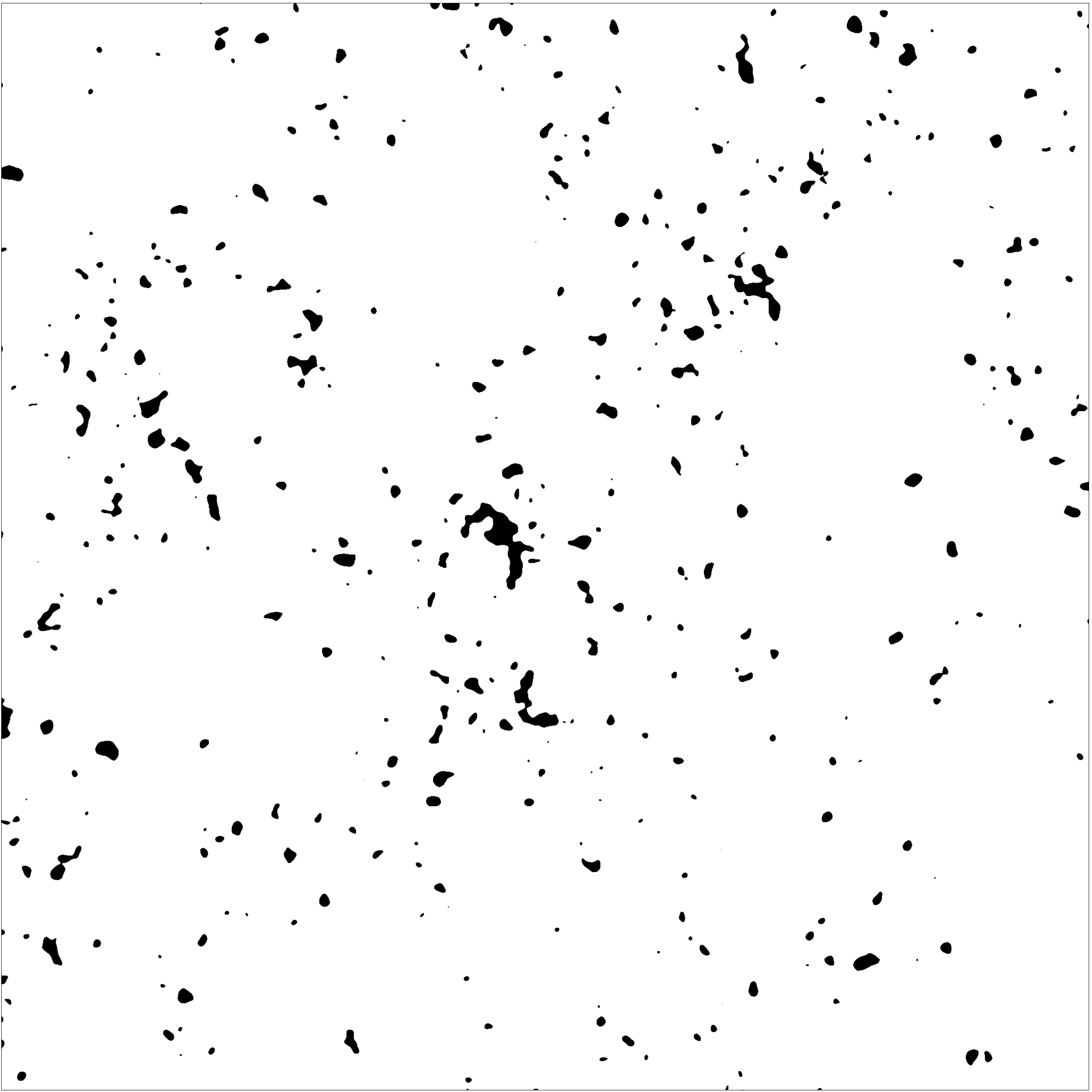}\\
\includegraphics[width=8 cm]{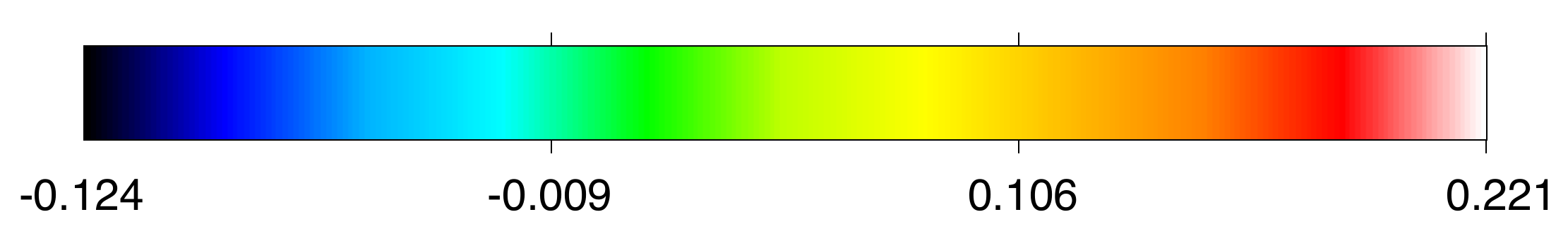}
\hfill
\caption[]{\textit{Top left panel: example of a simulated
    12-square-degree convergence map in the fiducial cosmology, with
    intrinsic ellipticity noise from source galaxies and $\theta_G=1$~arcmin
    Gaussian smoothing. A source galaxy density of $n_{gal}=15$/arcmin$^2$ at
    redshift $z_s=2$ was assumed. Other three panels: the excursion sets above
    three different convergence thresholds $\kappa$, i.e.\ all pixels
    with values above (below) the threshold are black (white). The
    threshold values are $\kappa=0.0$ (top right), $\kappa=0.02$
    (bottom left), and $\kappa=0.07$ (bottom right). The Minkowski
    Functionals $V_0$, $V_1$, and $V_2$ measure the area, boundary
    length, and Euler characteristic (or genus), respectively, of the
    black regions as a function of threshold. }}\label{fig:slice}
\end{figure*}

\begin{figure*}[htp]
\centering
\includegraphics[width=8 cm]{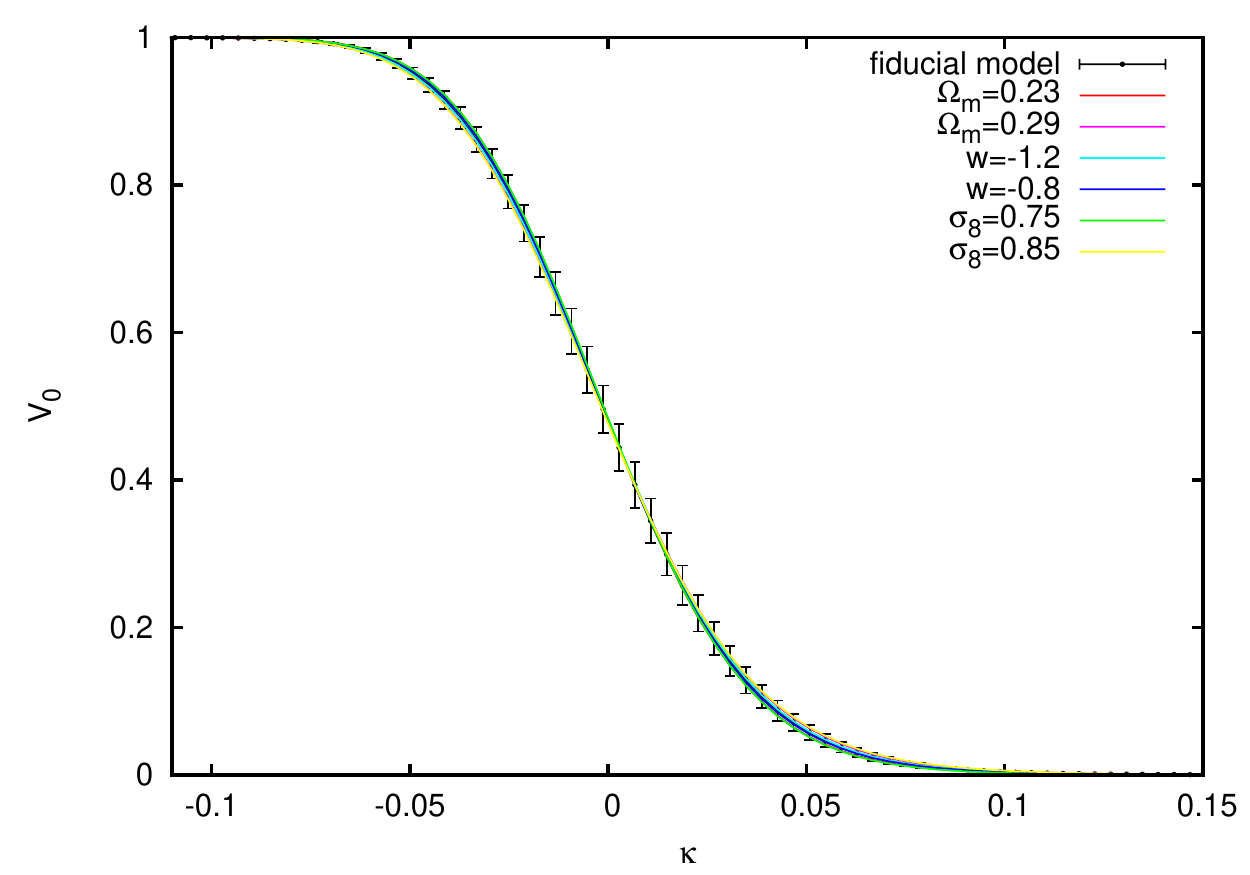} \includegraphics[width=8 cm]{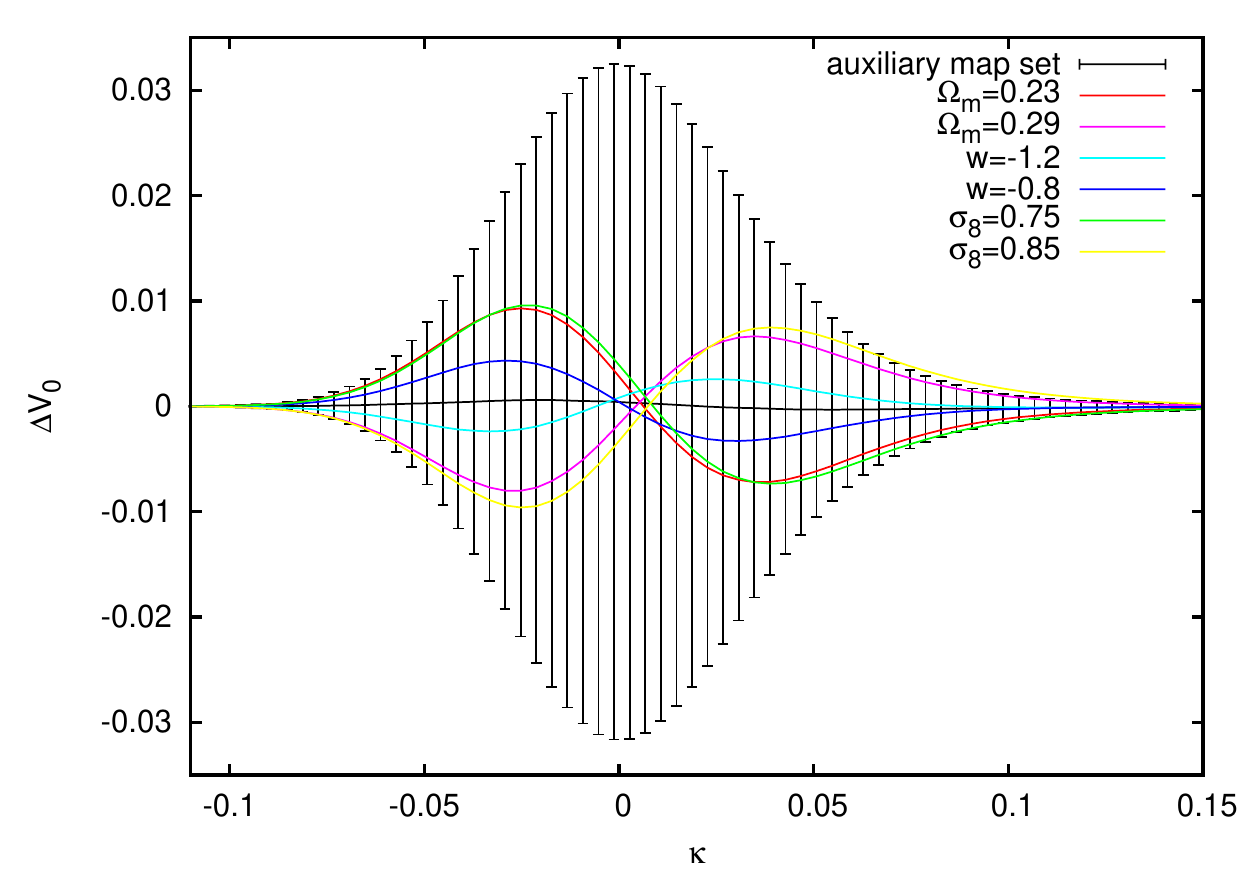} \\
\includegraphics[width=8 cm]{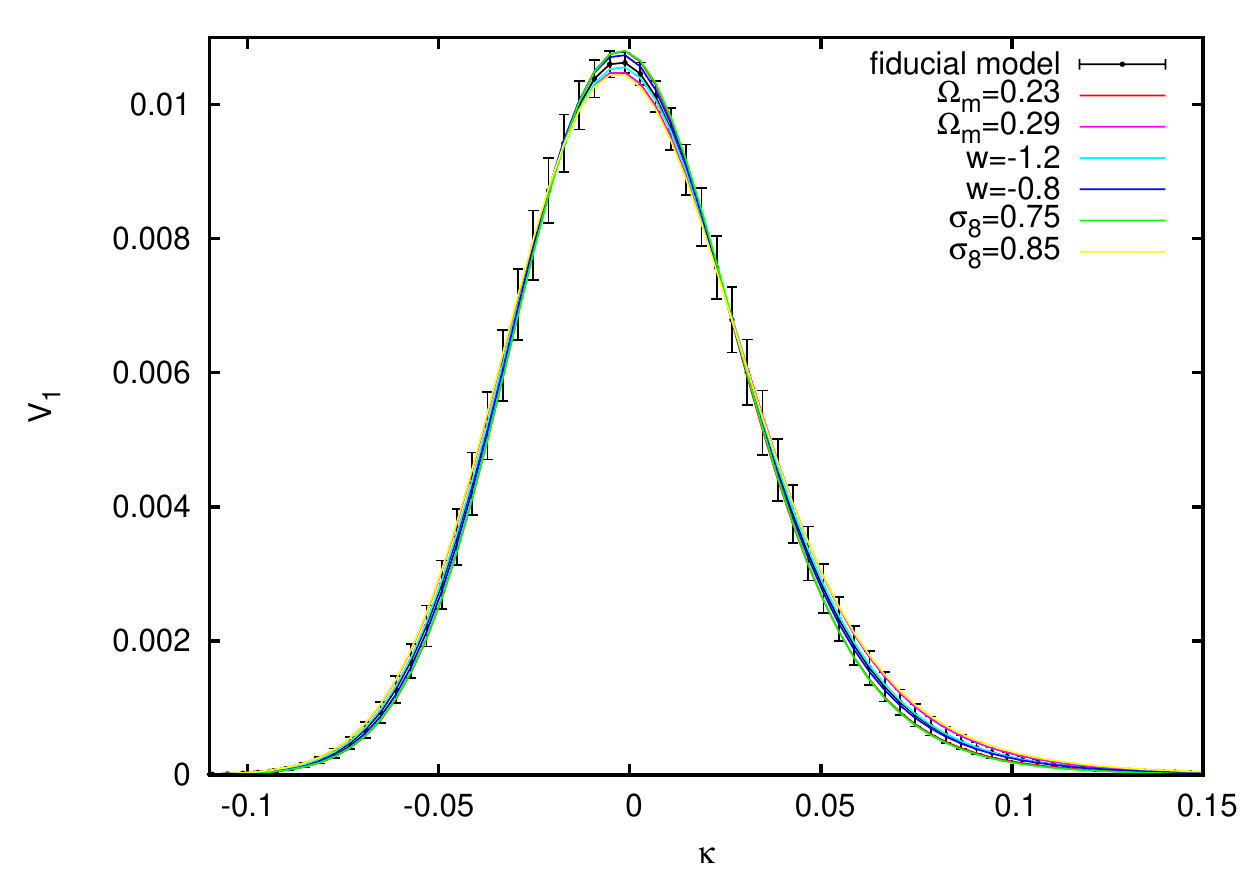} \includegraphics[width=8 cm]{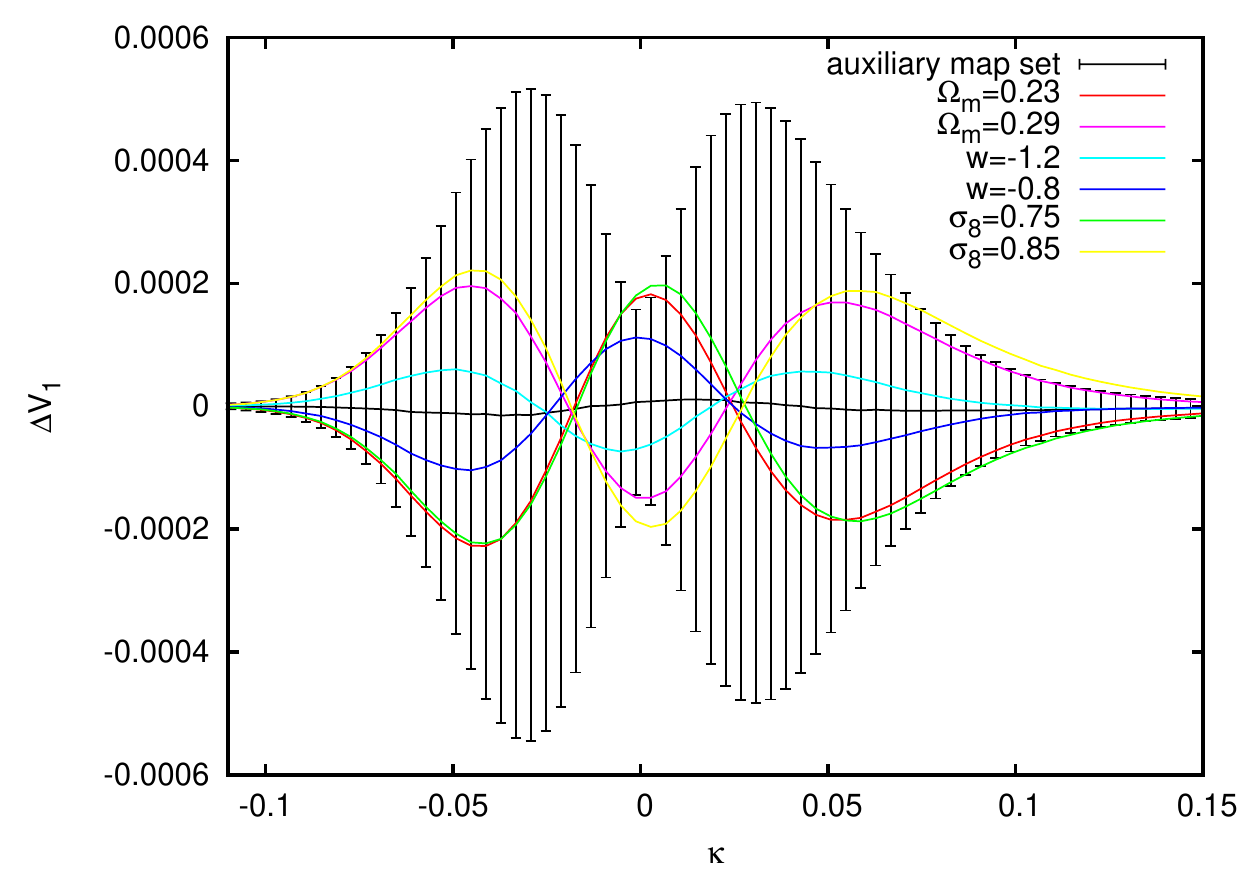} \\
\includegraphics[width=8 cm]{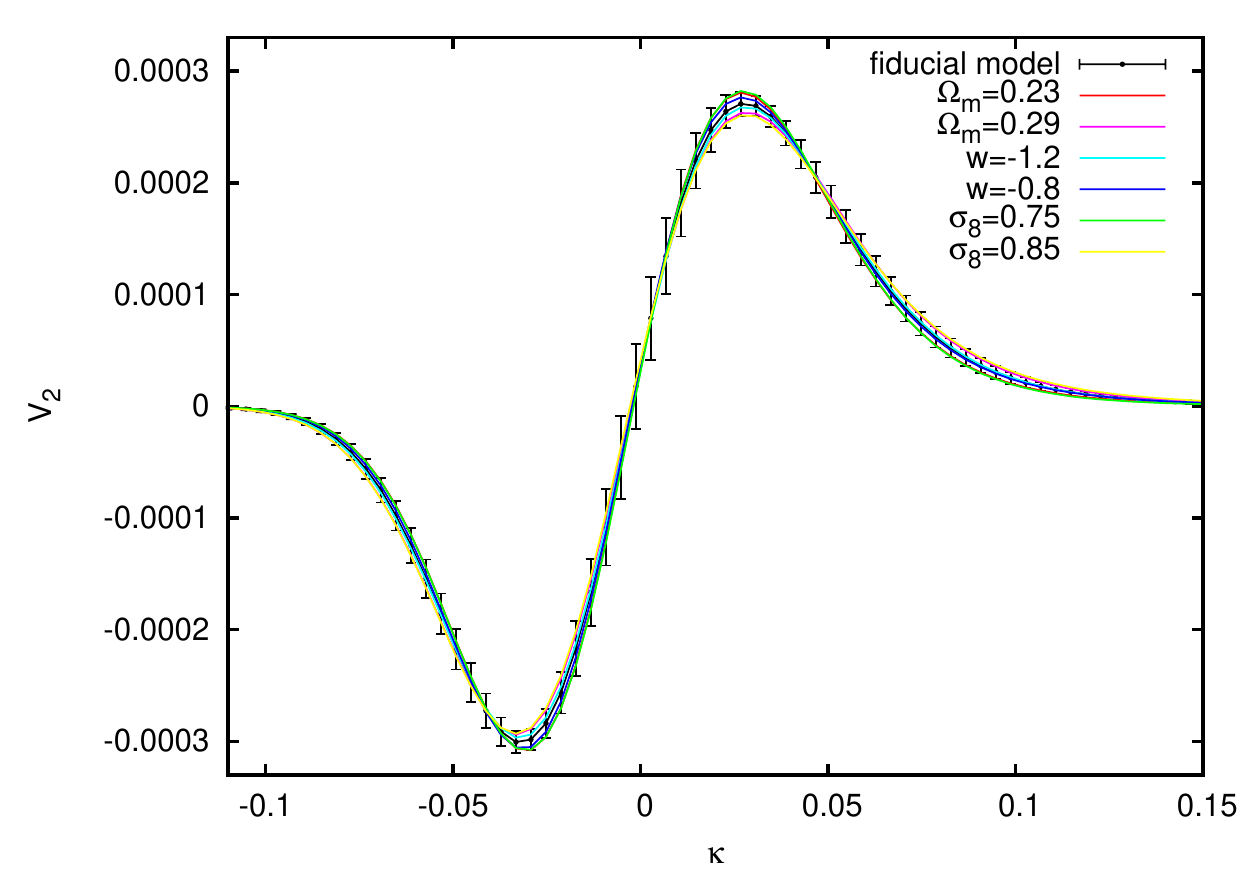} \includegraphics[width=8 cm]{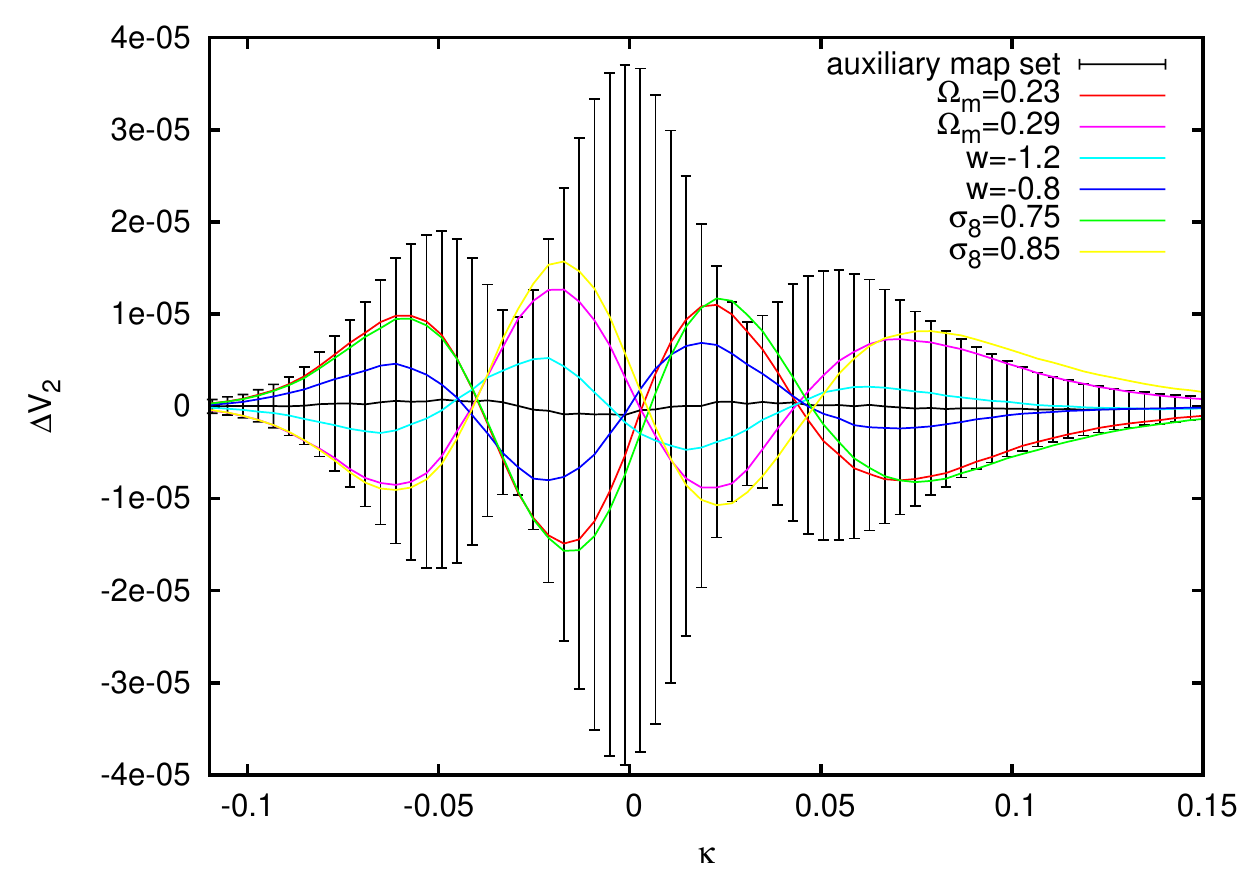}
\hfill
\caption[]{\textit{Left column: the three Minkowski Functionals in two
    dimensions, $V_0$ (area), $V_1$ (boundary length), and $V_2$
    (Euler characteristic) as a function of threshold $\kappa$ derived
    from 12-square-degree convergence maps in different cosmologies.
    Right column: differences of MFs in various cosmologies, compared
    to the fiducial model: control (``auxiliary'') map set in the
    fiducial model itself, with different realization of initial
    conditions (black), $\Omega_m=0.23$ (red), $\Omega_m=0.29$ (pink),
    $w=-0.8$ (blue), $w=-1.2$ (turquoise), $\sigma_8=0.75$ (green),
    and $\sigma_8=0.85$ (yellow).  The black error bars show the
    standard deviation of the MFs among the 1,000 maps of the fiducial
    map set. A source galaxy density of $n_{gal}=15$/arcmin$^2$ at
    redshift $z_s=2$ was assumed and $\theta_G=1$~arcmin Gaussian
    smoothing was applied.  }}\label{fig:Minkowski}
\end{figure*}

For illustrative purposes, Figure~\ref{fig:slice} shows one of our
12-square-degree convergence maps in the top left panel, and the
excursion set (shown as the black regions) in the other three panels
for the three different thresholds $\kappa=0.0, 0.02$ and $0.07$.  The
left column of Figure~\ref{fig:Minkowski} further shows the mean MFs
in each of the 7 cosmological models studied in this paper, averaged
over all 1,000 maps in each case ($V_0$, $V_1$, and $V_2$; top to
bottom).  The right column shows the difference between the mean MF in
the fiducial cosmology and the corresponding MF in each of the other
cosmologies. The error bars in all panels denote the standard
deviation among the 12-square-degree maps in the fiducial model (they
are similar in the other cosmologies and are omitted for clarity).

Since MFs are statistics on smoothed fields, while our convergence
maps are pixelized, one might worry that a discretized implementation
of Eqs.~(\ref{eqn:V0})--(\ref{eqn:V2}) can lead to spurious
``residuals'' as seen by some earlier work on MF (see for example
Refs. \cite{Hikage:2006fe,Hikage:2008gy,Matsubara:2010te}).  It turns
out the residuals are in fact \emph{not} caused by pixelization, but
instead by the discretization of the thresholds, i.e. finite
bin-sizes, as first pointed out by Ref. \cite{Lim:2011kd}. These
errors scale like $(\delta \nu)^2/\sigma_0$ which can be analytically
calculated and subtracted if the underlying distribution is known. On
the other hand, if we do not know the underlying map, we can work
around this problem by using small bin-sizes. While a small bin size
means that each bin is more noisy as there are less pixels binned,
there are more bins hence once integrated over the amount of
information remains the same.

To check for this, we generated 200 $2048\times 2048$ pixelized maps
of Gaussian random fields (GRF), and numerically calculated the
MFs. In the GRF case, the expectation values for the MFs can be
calculated analytically \cite{Matsubara:2003yt},
\begin{equation}\label{V0_GRF}
  V_0^{\mathrm{GRF}}(\nu) = \frac{1}{2}\left[1-\mathrm{erf}\left(\frac{\nu-\mu}{\sqrt{2}\sigma_0}\right)\right],
\label{eqn:GRFV0}
\end{equation}
\begin{equation}\label{V1_GRF}
V_1^{\mathrm{GRF}}(\nu) = \frac{1}{8\sqrt{2}}\frac{\sigma_1}{\sigma_0}\exp\left(-\frac{(\nu-\mu)^2}{2\sigma_0^2}\right),
\label{eqn:GRFV1}
\end{equation}
and
\begin{equation}\label{V2_GRF}
V_2^{\mathrm{GRF}}(\nu) = \frac{\nu-\mu}{4\sqrt{2}}\frac{\sigma_1^2}{\sigma_0^3}\exp\left(-\frac{(\nu-\mu)^2}{2\sigma_0^2}\right),
 \label{eqn:GRFV2}
\end{equation}
where $\mu=\langle\kappa\rangle$ is the mean,  
\begin{equation}\label{sigma0}
\sigma_0 = \sqrt{\langle \kappa^2\rangle - \mu^2}
\end{equation}
is the standard deviation, and
\begin{equation}\label{sigma1}
\sigma_1 = \sqrt{\langle\kappa_x^2+\kappa_y^2\rangle}
\end{equation}
is its first moment.

\begin{figure}[htp]
\centering
\includegraphics[width=7 cm]{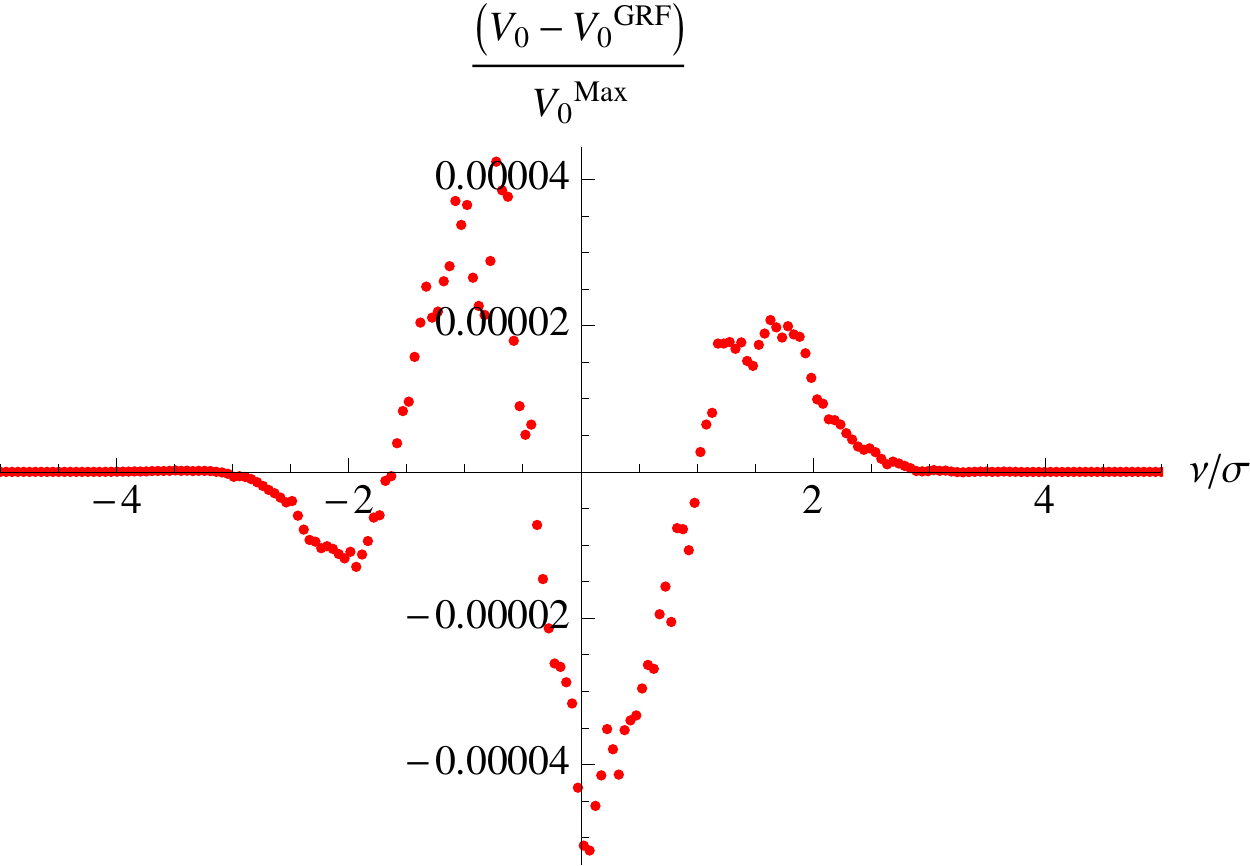} 
\includegraphics[width=7 cm]{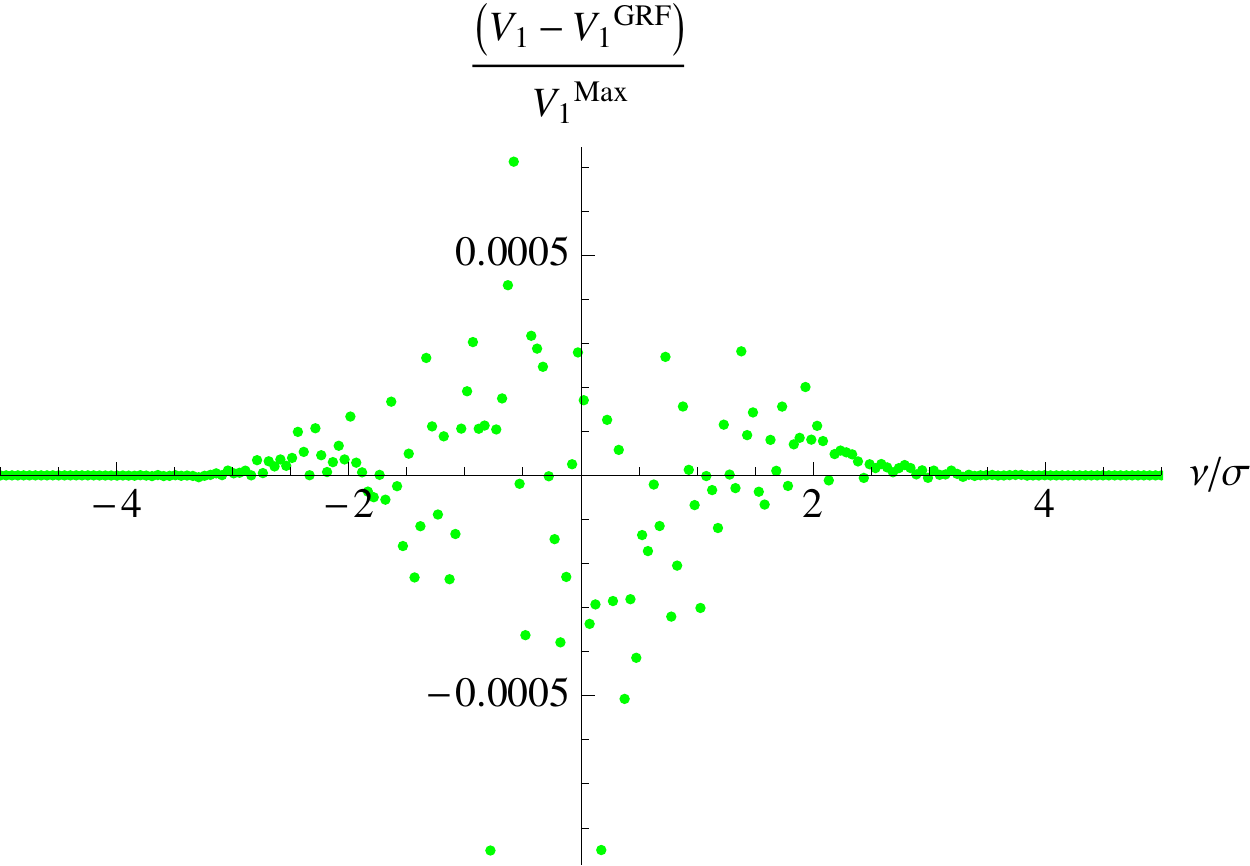} 
\includegraphics[width=7 cm]{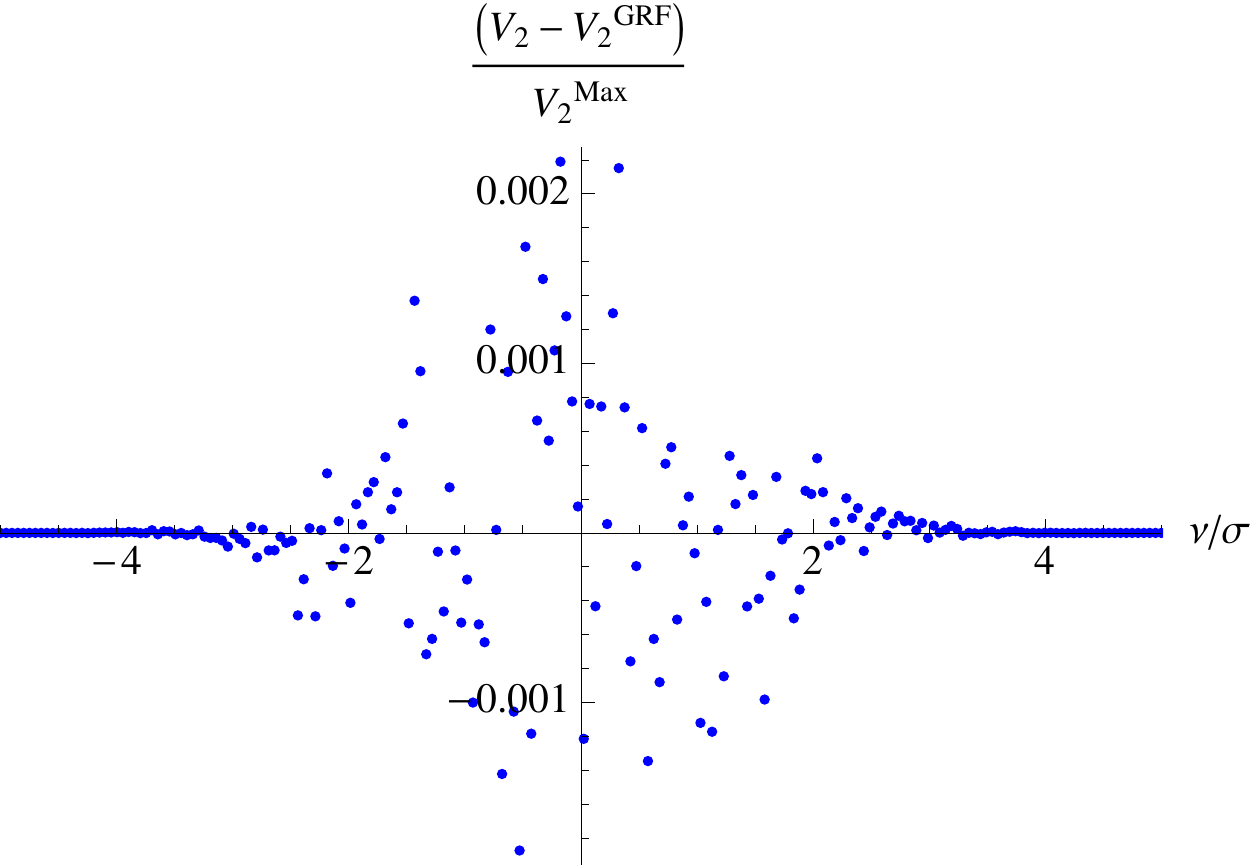}
\hfill
\caption{Comparison between the average MFs numerically calculated
  from 200 mock Gaussian random field maps with unit variance, and the
  analytic expectations given by
  Eqns.~(\ref{eqn:GRFV0})--(\ref{eqn:GRFV2}).  The x axis shows the
  value of the threshold, and the y axis shows the fractional
  differences in the MFs normalized to the maximum value of the
  respective MF, as labeled.}
   \label{fig:dGRF}
\end{figure}

We average the MFs measured from these 200 GRF maps, each with 200
threshold bins from $\nu/\sigma_0=-5$ to $\nu/\sigma_0=5$, to find the
mean $\langle V_j\rangle$, and then compare it to the analytic
expressions given by Eqns. (\ref{eqn:GRFV0})-(\ref{eqn:GRFV2}).  In
the latter, we use the values $\langle \mu \rangle$, $\langle
\sigma_0\rangle$ and $\langle \sigma_1 \rangle$ obtained by averaging
over the same 200 maps. As can be seen from Fig.~(\ref{fig:dGRF}),
which shows the difference between our numerically calculated and
theoretically expected MFs, our procedure reproduces the MFs in the
GRF case highly accurately.

It is instructive to briefly comment on what the MFs measure in
broader, qualitative terms. The properties of random fields can be
roughly divided into four distinct categories: (a) the histogram,
which probes the distribution of pixel values, but is insensitive to
their spatial distribution, (b) spatial correlations, which depend on
the distances between pixel values, but not about the shapes of
structures that lie in between, (c) the shape of objects, and (d)
topology, which cares about the connectedness of objects, but by
itself is insensitive to their distances or shapes, as it is
independent of the underlying metric.

The power spectrum is a pure measure of (b), while the MFs are also
sensitive to (a), (c) and (d). In fact, $V_0$, the cumulative
probability distribution function (PDF), is a pure measure of
(a). $V_2$, the Euler characteristic of the excursion set, measures
the topology (d). However, $V_2$ is not a pure measure of
connectedness of structures, as it is \lq\lq contaminated\rq\rq\ by
the histogram. Intuitively, $V_1$ also contains information on
(c). This broad classification of information will be useful when
interpreting the origin of the constraints from the MFs for different
smoothing scales.

\subsection{Power Spectra}
\label{Power Spectra}

In order to study the non-Gaussian information content of the MFs
explicitly, we also compute the power spectrum from our convergence
maps.  The power spectra were first pre--computed for 1000 equally
spaced bins of the angular wave vector $\vec\ell$ between $100 \leq
|\vec\ell|\leq 100,000$, covering the full range of angles from our
pixel size ($\sim 6$ arc-sec) to the linear size of our maps ($\sim
3.5$ deg).  We further adopted the flat sky approximation, assumed
spatial isotropy, and averaged modes with the same length of the wave
vector $\vec\ell$ in different directions, to find the 1D power
spectrum $P(\ell)$ as a function of $\ell=|\vec\ell|$ alone.

\begin{figure}[htbp]
\centering
\includegraphics[width=8 cm]{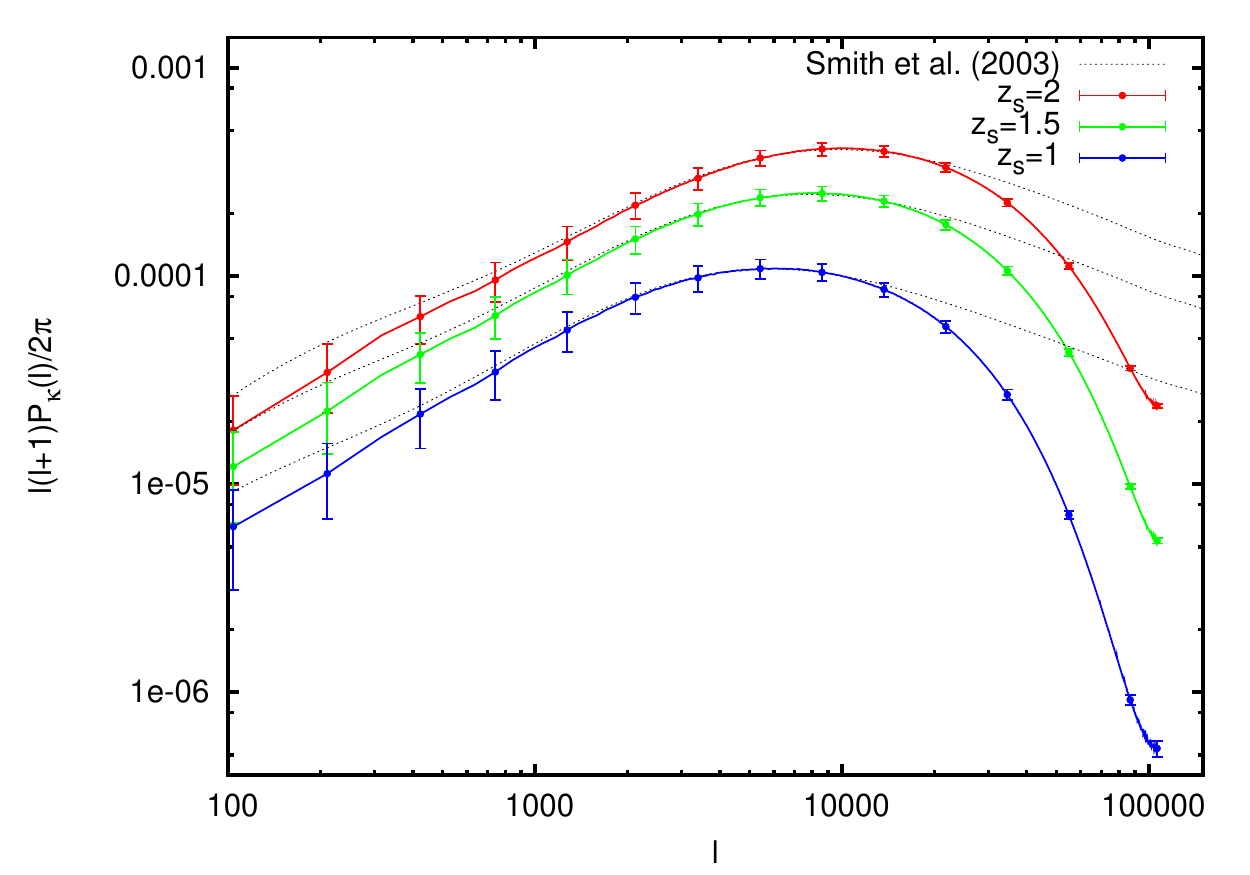}  
\hfill
\caption[]{\textit{Convergence power spectra in the fiducial model
    from our simulations, for source galaxies at redshift $z_s=1,
    1.5$, and $2$ (from bottom to top, in blue, green, and red,
    respectively) compared to theoretical predictions calculated with
    Nicaea \cite{Nicaea} (black dashed curves). The simulated spectra
    have been averaged over 1,000 maps, without ellipticity noise or
    smoothing, and the error bars show the r.m.s.\ variation among
    these maps. Power is missing on large scales due to the finite box
    size of the simulations, and on small scales due to mass
    resolution and resolution on the $4096^2$-pixel density planes.
    }}\label{fig:power spectrum}
\end{figure}

The power spectrum can be derived using the Limber approximation
\cite{Limber} and integrating the 3D matter power spectrum along the
line of sight.  To check the accuracy of our simulations, ray-tracing
code, and construction of the lensing maps, we compare our numerically
measured lensing power spectra to the fitting formulae in
\cite{Smith}.  For this comparison, we derived power spectra from raw
12-square-degree WL maps (without adding noise or smoothing), and
averaged over the 1,000 maps in our fiducial model.  The results are
shown by the solid curves in Figure~\ref{fig:power spectrum} for the
three different redshifts $z_s=1, 1.5$, and $2$, with error bars
showing the standard deviation of the power in each $\ell$-bin. The
dashed curves show the expectations from \cite{Smith}, which we
computed with the public code Nicaea \cite{Nicaea}.

The figure shows that our simulations lose power below $\ell\sim400$,
due to our finite box size. On smaller scales, we find excellent
agreement out to $\ell\sim20,000$ for $z_s=1$ and out to
$\ell\sim30,000$ for $z_s=1,5$ and $2$, corresponding to our
resolution limit.  Because of this limitation, we will employ
smoothing scales no smaller than 1 arcmin below.  Comparing
Figure~\ref{fig:power spectrum} to Figure~3 in
\cite{Kratochvil:2009wh}, we notice that the drop-off in power has
been pushed out to higher $\ell$, due to the increased resolution of
the density planes.

\begin{figure}[htbp]
\centering
\includegraphics[width=8.3cm]{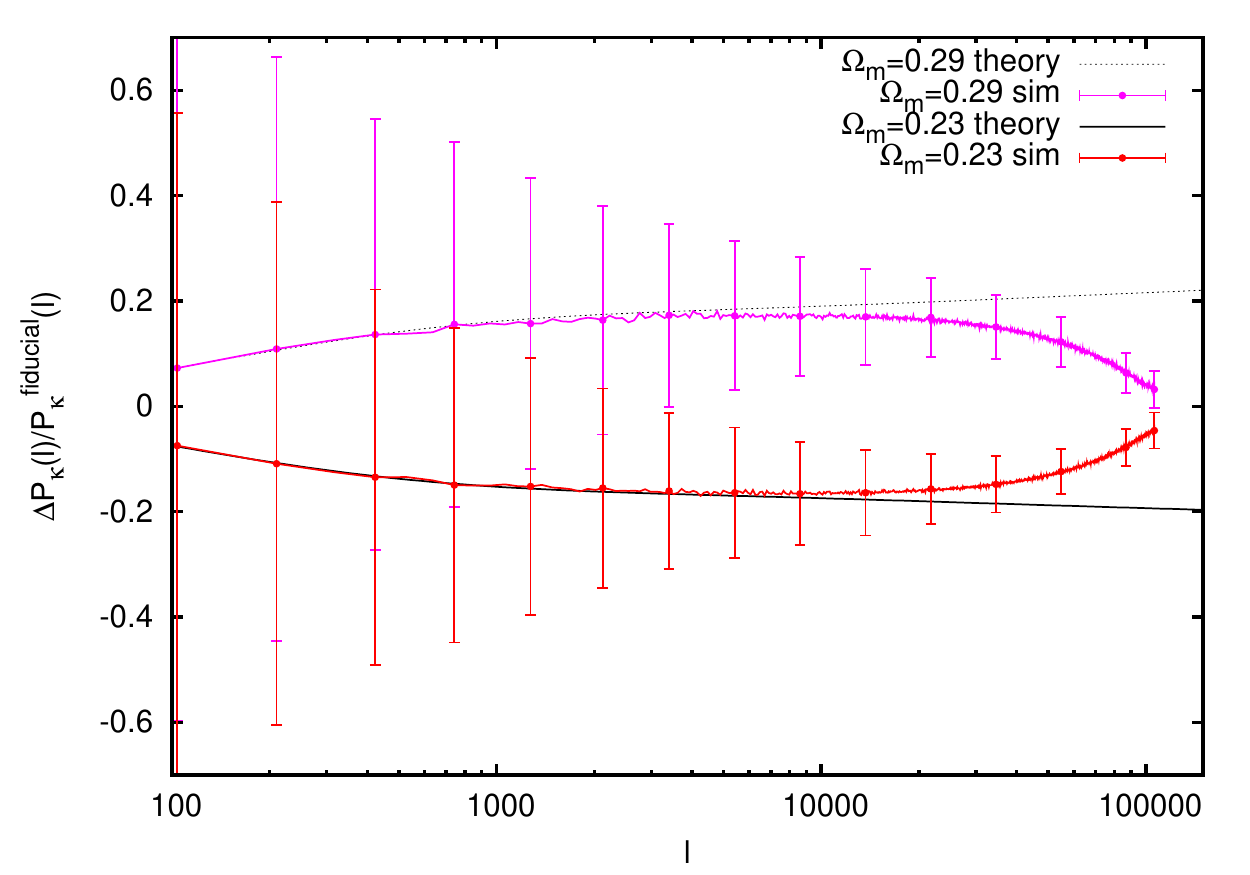}\\
\includegraphics[width=8.3cm]{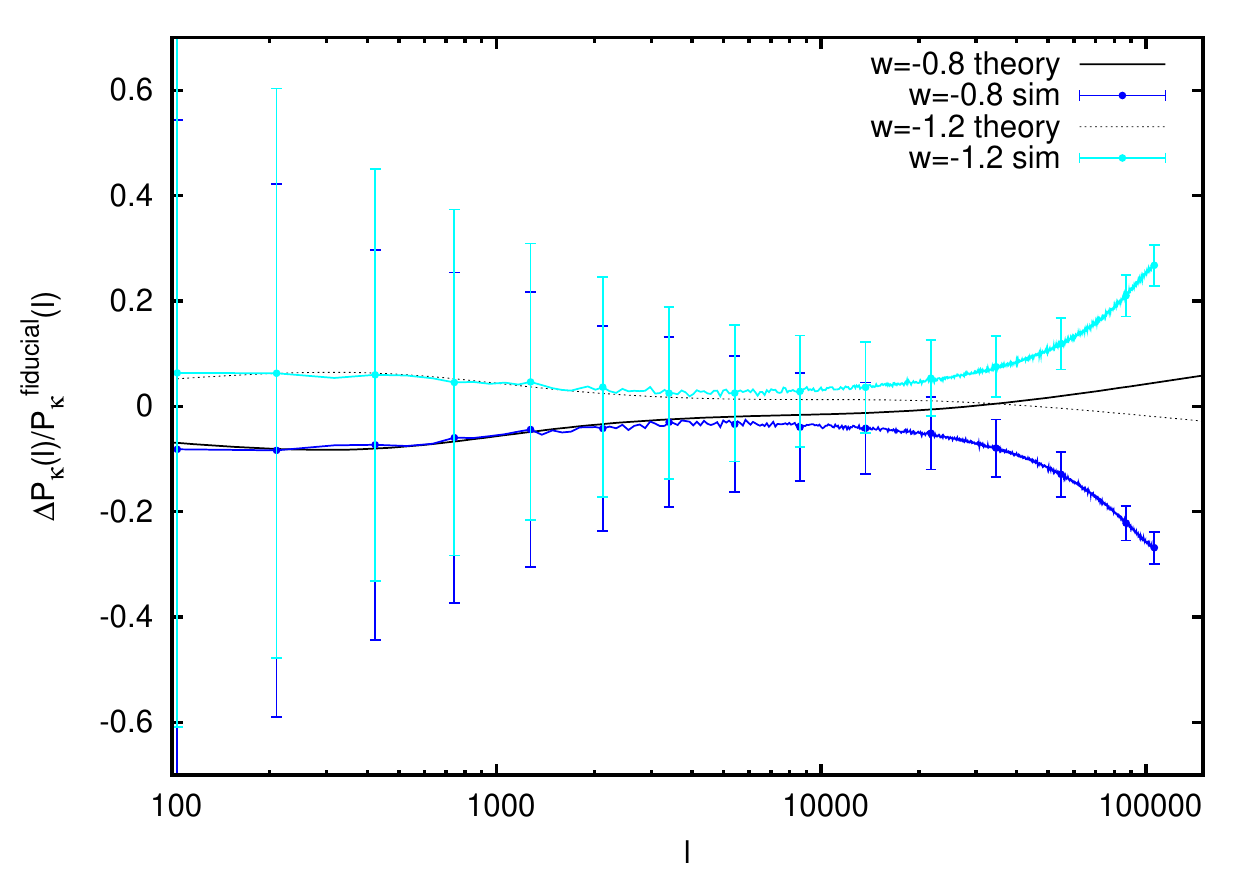}\\
\includegraphics[width=8.3cm]{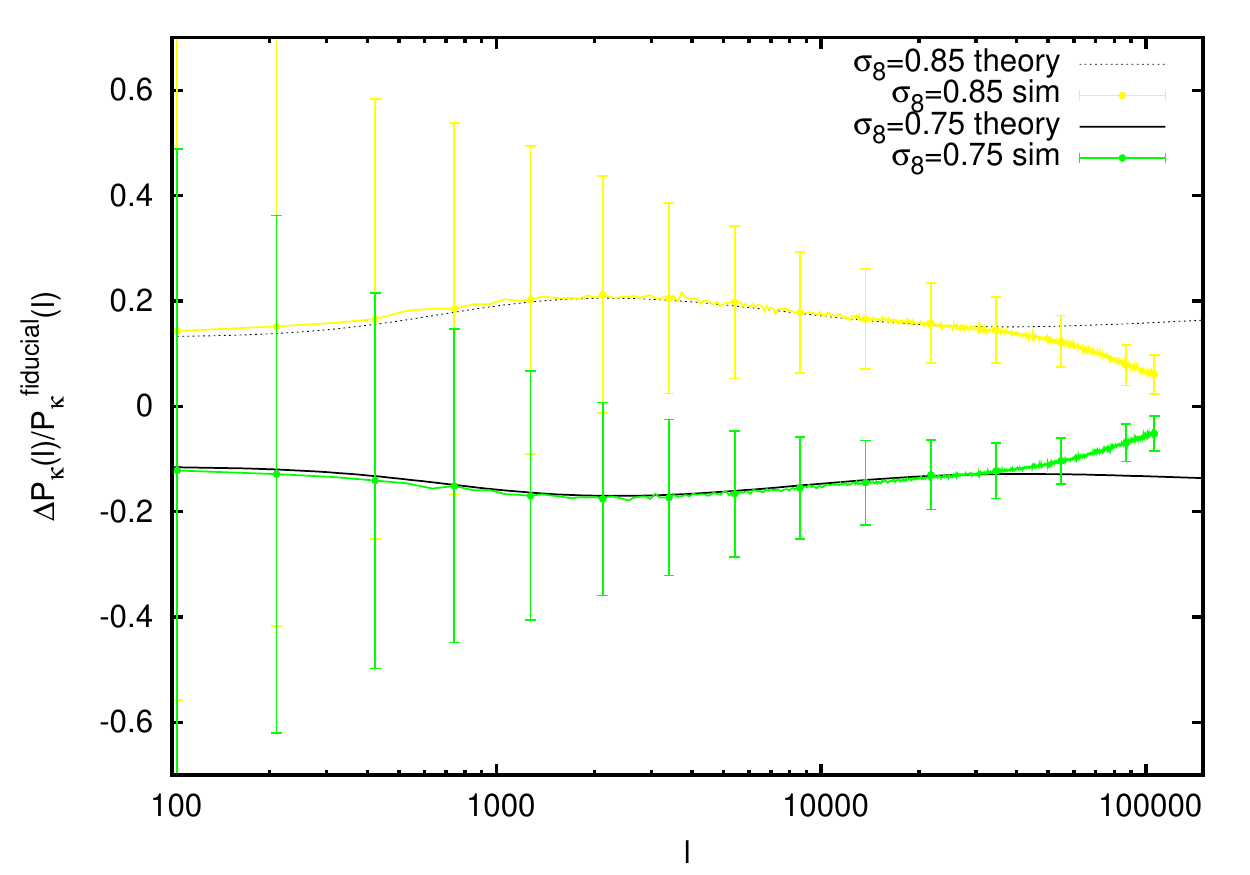}
\hfill
\caption[]{\textit{Convergence power spectra derived from our
    simulations are compared to theoretical predictions, as in
    Figure~\ref{fig:power spectrum}, except we only show results for
    source galaxies at redshift $z_s=2$, and show the {\em
    differences} between the power spectra in various cosmologies
    compared to the fiducial model.  }}\label{fig:power spectrum diff}
\end{figure}

Our results rely mostly on the cosmology-dependence of the power
spectrum (and MFs), rather than its absolute value. We therefore
compare the {\em differences} of the power spectra in various
cosmologies from the fiducial case. The results are shown in
Figure~\ref{fig:power spectrum diff}, which shows that the agreement
is excellent for the dependence of the power spectrum on all three
parameters, down to scales of $\ell\sim 20,000$.  The theoretical
predictions for the underlying non-linear matter power are good only
to about $\sim$10\% on scales down to $\sim 0.1$Mpc~\cite{CoyoteI},
which can cause an under-prediction for the convergence power spectrum
as well \cite{Hilbert:2008kb, Sato:2009, Eifler:2010kt}; the
differences are therefore within the accuracy of the theory down to
these $\ell$.

%%%%%%%%%%%%%%%%%%%%%%%%%%%%%%%%%%%%%%%%%%%%%%%%%%%%%%%%%%%%%%%%%%%%%%%%%%%%%%%%%%%%%%%%%%
\section{Statistical Methods of Analysis}
\label{Statistics}
%%%%%%%%%%%%%%%%%%%%%%%%%%%%%%%%%%%%%%%%%%%%%%%%%%%%%%%%%%%%%%%%%%%%%%%%%%%%%%%%%%%%%%%%%%

\subsection{Statistical Descriptors}

Together, we generically refer to the MFs and the power spectra as
statistical ``descriptors'' of the convergence map.  The descriptors
can be combined into a single vector, $N_i$, where $i$ indexes the
threshold for MFs and the multipole for the the power spectrum.
Combining the data from several source redshifts or smoothing scales
simply makes the descriptor vector longer, but we still treat it in
the same way.  For each MF, we divide the range into 15 threshold
bins.  Similarly, we divide the power spectrum into 15 scale bins (we
require such a small number to keep the covariance matrix tractable,
as discussed below). The most comprehensive case we consider uses
three source redshifts, each with five smoothing scales for each of
the three MFs, plus the power spectrum, for a total of
$15\times3\times5\times4 =900$ entries in the $N_i$ vector.

To constrain cosmology, we are interested in the true ensemble average\footnote{Averaged
over all possible universes with the same cosmological parameter values.}
(denoted henceforth by brackets $\langle\ \rangle$) and covariance of
these descriptors as a function of cosmological parameters
($\mathbf{p}=\{\Omega_m, w, \sigma_8\}$).  These of course are not
available to us, but can be estimated from the simulations.  Averaging
over the pseudo-independent map realizations within a given cosmology,
we can estimate the ensemble average by 
\BE\label{SM}
\langle{N}_i(\mathbf{p})\rangle\approx\overline{N}_i(\mathbf{p})\equiv\frac{1}{R}\sum_{r=1}^R
N_i(r, \mathbf{p}),
\EE
where $N_i(r, \mathbf{p})$ is the descriptor vector for a single
realization and $r$ runs over our $R=1000$ map realizations.  We call
this estimate the \textit{simulation mean}.  It differs from the true
ensemble average both because of the limited number of realizations
and also because of the limitations inherent in our simulations.  In
the absence of a fitting formula for the MFs in the non-Gaussian case
(analogous to the power spectrum formula from \cite{Smith}) the
simulation mean serves as our proxy for theoretically predicted
MFs.~\footnote{Alternatively, the recently proposed perturbative
description of WL MFs \cite{Munshi:2011wu} could serve this purpose.}

Because of the computational expense, we can only form this estimate
at our few selected cosmologies (Table~\ref{tab:Cosmologies}).  Using
finite differences between the simulated cosmologies, we construct a
first-order Taylor expansion around our fiducial cosmology to estimate
$\overline{N}_i(\mathbf{p})$ for other cosmologies not explicitly
covered by simulations:
\BE\label{Taylor}
\overline{N}_i(\mathbf{p})\approx\overline{N}_i(\mathbf{p_0})+\sum_\alpha \frac{\overline{N}_i(\mathbf{p}^{(\alpha)})-\overline{N}_i(\mathbf{p_0})}{p^{(\alpha)}_\alpha-p_{\mathbf{0}\alpha}}\cdot(p_\alpha-p_{\mathbf{0}\alpha}),
\EE
Here, index $\alpha=1,2,3$ refers to an individual parameter, such as
$\Omega_m, w$, or $\sigma_8$, and the sum counts through all varied
cosmological parameters, while $\mathbf{p}^{(\alpha)}$ denotes the
cosmological parameter vector of an actually simulated non-fiducial
cosmology with only the parameter $p_\alpha$ varied. The fraction in
Eq.~(\ref{Taylor}) is the finite difference derivative. If the
non-fiducial cosmology is chosen such that
$p^{(\alpha)}_\alpha-p_{\mathbf{0}\alpha}$ is positive, we call it a
``forward derivative'', if it is negative, we call it a ``backward
derivative''. We use either one or the other derivative to compute
parameter constraints to assess the robustness of our results.

Similarly to the simulation mean, we estimate the covariance of the
statistical descriptors from the simulations, ${\rm Cov}( N_i, N_j)
\approx C_{ij}$, where
\BE
\label{covariance matrix}
C_{ij}(\mathbf{p})\equiv\frac{1}{R-1}\sum_{r=1}^R [N_i(r,\mathbf{p})-\overline{N}_i(\mathbf{p})][N_j(r,\mathbf{p})-\overline{N}_j(\mathbf{p})].
\EE
This covariance matrix contains contributions both from the sample
variance of the true convergence signal and from the random
ellipticity noise.  When the size of this covariance matrix is large,
inaccuracies in its estimate can become challenging, as we explore
further below.

\subsection{Parameter Estimation and Constraints}
\label{Monte Carlo}

In principle -- having numerically evaluated Eqs.~(\ref{SM}),
(\ref{covariance matrix}) and the finite difference derivatives for
Eq.~(\ref{Taylor}) from the simulated WL maps -- one can use the
Fisher matrix formalism\footnote{See e.g.~Eq.~15 in
  \cite{Tegmark:1996bz} and Eq.~62 in \cite{Heavens:2009nx}.} to
compute parameter constraints.  In fact, at least two groups have
followed this approach recently for weak lensing simulations with
redshift tomography \cite{Yang:2011zz, Seo2011}.  In practice,
however, the Fisher matrix is a forecasting tool and mis-estimation of
the covariance from the (simulated) data makes this procedure unstable
as we combine smoothing scales and redshifts.  As the descriptor $N_i$
surpasses several hundred entries, the marginalized error on
parameters $\Delta p_\alpha^{\rm
  marg.}=\sqrt{(F^{-1})_{\alpha\alpha}}$ derived from the Fisher
matrix $F$ shows a runaway behavior to smaller errors.  (Here again,
index $\alpha=1,2,3$ refers to an individual parameter, such as
$\Omega_m, w$, or $\sigma_8$.)  The explanation for this behavior is
straightforward, and relates to the imprecise estimation of the
covariance matrix from a small number of independent realizations:
When the number is insufficient, outliers will be absent, so
covariances will typically be underestimated, resulting in overly
optimistic error-bars.

Since we want to combine many descriptors, we need a procedure which
is robust and conservative when the covariance is poorly estimated.
With these requirements, we chose to use $\chi^2$-minimization to fit
for parameters, then measure the distribution of parameter fits using
an ensemble of Monte Carlo realizations.

For realizations drawn from the fiducial cosmology $\mathbf{p_0}$,
$\chi^2$ is
\BE 
\label{chi2}
\chi^2(r, \mathbf{p})\equiv\sum_{i,j}\, \Delta N_i(r, \mathbf{p})\,[{\rm Cov}^{-1}\mathbf{(p_0)}]_{ij}\, \Delta N_j(r,\mathbf{p})
\EE
where
\BE
\Delta N_i(r,\mathbf{p})\equiv N_i(r,\mathbf{p_0})- \langle{N}_i(\mathbf{p})\rangle.
\EE
For each Monte Carlo realization, we minimize $\chi^2$ with respect to
$\mathbf{p}$ using a simulated annealing algorithm.  In practice, the
simulation-based estimates described above replace the ensemble
average and covariance.  Note that we fix the covariance matrix to our
estimate at the fiducial cosmology, $C_{ij}(\mathbf{p_0})$, but below
argue that, for this procedure, having the exact covariance matrix is
not crucial.  The covariance matrix is inverted with singular value
decomposition and condition number $10^6$, discarding any problematic
eigenvectors.  The simulation means, finite differences between
cosmologies, and the covariance matrices are all computed with the
auxiliary data set (5 N-body simulations), which has the same
cosmology as the fiducial data set, but shares random seeds, and
therefore large-scale structures, with the realizations from the
alternative cosmology simulations.  In this way we probe more directly
the effect of cosmological differences on these statistical
descriptors.  It is also important that the covariance matrix is
constructed from a map set that is strictly independent of the maps
for which $\chi^2$ is being evaluated, lest the covariance be
erroneously ``tuned'' to the specific data set.

The maps built from the much larger fiducial data set (45 N-body
simulations) make up our Monte Carlo ensemble.  The $N_i$ vectors
computed from these maps are our best representation of the
distribution of measured descriptors.  The distribution of the
parameters fit to these descriptors is used to compute the error bars
and confidence contours below.  Marginalized errors are computed from
the variance of each parameter.  To illustrate covariances, we plot
approximate two-dimensional error ellipses by computing the covariance
between parameters (evaluated over the Monte Carlo ensemble), drawing
the 68.4\% confidence limit from a corresponding bivariate Gaussian
distribution with the same covariance.  We checked these contours
against the Fisher matrix contours in a few cases, finding good
agreement when the covariance matrix is small.

For a large numbers of bins, the Monte Carlo confidence limits are
stable, and do not show the runaway behavior we saw for the Fisher
matrix estimates.  Furthermore, the parameter fits are not strongly
biased by a bad choice of covariance matrix (if the system were linear
there would be zero bias) although the resulting error bars are not
optimal.  This means that errors in the estimate of the covariance
matrix tend to make our estimate more conservative.
Finally, we introduce the notation
\BE
\chi^2_{\rm min}(r)\equiv\min_\mathbf{p} \chi^2(r,\mathbf{p})
\EE
for the minimized $\chi^2$ for map $r$, and $\chi^2_{\rm min}/n$ for
the $\chi^2$ per degree of freedom in this best-fit model. Here the
number of degrees of freedom is the length of $N_i$ minus the three
model parameters varied.  Averaged over realizations,
$\overline{\chi}_{\rm min}^2/n$ is an indicator for the typical
goodness of fit (and in the case of Gaussian errors, would be
$\lesssim1$).  On the other hand, errors in the covariance estimate
will make the fit worse, so we use this quantity as a diagnostic of
the quality of the estimate.

\subsection{Binning}
\label{Binning}

Here we give more details on our choices of binning. We pre-compute
the MFs for 200 equally spaced convergence thresholds $\kappa$ and the
power spectrum for 1,000 equally spaced values of $\ell$.  Before the
analysis, we divide the MFs into 15 equally spaced bins in
$\kappa$. To avoid bins lying in noisy regions where outliers in a few
maps dominate (i.e. very high and low thresholds $\kappa$), we
restrict ourselves to the range where the simulation mean of the MFs
is at least 5\% of its maximum value. In the case of $V_0$, we use the
differential (instead of the standard cumulative) version of the MF to
determine the $\kappa$ range.

The power spectrum is originally computed in narrow, equally spaced
bins, which we rebin into 15 logarithmic bins (based on the bin
minimum), as close to logarithmic as possible without interpolation.
The logarithmic bins span $\ell = 100$--$20,000$, from the largest
mode in the map to our smallest smoothing scale $\theta_G=1$ arcmin.
Using upper cutoffs at $\ell=40,000$ and $\ell=80,000$ does not change
the results appreciably.

\subsection{Breakdown of the Covariance Matrix Estimate}
\label{Maximum Size of the Covariance Matrix}

The size of our covariance matrix varies from $15\times15$ elements
for a single descriptor, redshift, and smoothing scale to
$900\times900$ elements for all three MFs combined with the power
spectrum, redshift tomography with three source galaxy planes, and
five smoothing scales all combined.  Unless all descriptors are highly
correlated, it is clearly very difficult to estimate all elements
accurately from merely $R = 1000$ maps.  We therefore expect that as
we increase the number of threshold or multipole bins, combine more
descriptors, redshifts, or smoothing scales, at some point, our
results become unreliable just because we have too few maps.

When we compute the marginalized errors both from the Fisher matrix
and from the Monte Carlo procedure, as mentioned above, they agree for
small $N_i$ vectors. Ideally, as the vector's number of entries is
increased beyond the point where no more information can be gleaned
from the fine shape of the MFs or the power spectrum, the
errors computed from the Fisher matrix and from the Monte Carlo method
would both reach a plateau. In practice, as one increases the number
of entries further beyond the quality limit of the dataset, the
marginalized errors from the Fisher matrix start {\em improving}
further as long as one keeps adding bins, even if there is no new
information content in them. This is clearly unphysical (and comes
from underestimating the [co]variances, as pointed out above).  In
contrast, we find that the Monte Carlo errors start {\em increasing}
modestly in such a situation. This behavior of the Monte Carlo results
is also unphysical---the constraints on parameters cannot become worse
when no or more information is added.

The difference between the behavior of the Fisher matrix and the Monte
Carlo is crucial when applying these techniques to simulation data, as
opposed to doing forecasting from a predictive analytic theory. In
particular, the derived constraints become unreliable if the quality
of the dataset is insufficient (or just barely sufficient) to reach
the plateau, i.e. the covariance matrix gets corrupted before (or just
as) the best constraints are reached. Then no plateau can be
identified and the Fisher matrix constraints keep continuously
improving -- most importantly, there is no indication of when one
transitions into the unphysical breakdown regime. The Monte Carlo
method, on the other hand, provides a conservative estimate in this
case: as long as the quality of the data allows it, the parameter
constraints will improve, and when the quality limit of the data is
reached---not because there is no more information in principle, but
because there is insufficient data to give good enough estimates for
the covariance matrix---the constraints start to worsen again
modestly. In that case one can simply take the tightest Monte Carlo
constraints achievable to get the best constraints which the given
dataset allows. In our cases with multiple descriptors, redshifts, and
many smoothing scales, we have found it difficult to get a good enough
dataset to see a plateau.

In addition to seeing the constraints degrade, the minimum value of
$\chi^2$ can indicate when this breakdown occurs.  Empirically, we
note that as the confidence contours reach their achievable minimum
from our data, the $\overline{\chi}^2_{\rm min}/n$ starts to rise,
eventually reaching $\overline{\chi}^2_{\rm min}/n \approx 2$ which
signals the beginning of the unphysical breakdown.  For example, when
all three MFs are combined, with all three redshifts and all five
smoothing scales, we have 675 entries in $N_i$ and
$\overline{\chi}^2_{\rm min}/n\approx1.9$, at which point the contours
become a few percent larger than for just three smoothing scales.
Further adding the power spectrum to these constraints results in 900
entries and $\overline{\chi}^2_{\rm min}/n\approx2.8$, and the
contours begin to widen noticeably.  We therefore take this limit as
the breakdown of our simulated map set.

This behavior, which is the opposite of the Fisher matrix approach
that tends to {\em underestimate} the errors, makes our Monte Carlo
results conservative with respect to inaccuracies in the covariance
matrix.

%%%%%%%%%%%%%%%%%%%%%%%%%%%%%%%%%%%%%%%%%%%%%%%%%%%%%%%%%%%%%%%%%%%%%%%%%%%%%%%%%%%%%%%%%%
\section{Results}\label{Results}
%%%%%%%%%%%%%%%%%%%%%%%%%%%%%%%%%%%%%%%%%%%%%%%%%%%%%%%%%%%%%%%%%%%%%%%%%%%%%%%%%%%%%%%%%%

Utilizing the Monte Carlo procedure, we were able to obtain reliable
constraints for combinations of three source redshifts $z_s=1, 1.5, 2$
and five smoothing scales $\theta_G=1', 2', 3', 5', 10'$ for nearly
all combinations of the MFs and power spectra.

\begin{figure*}[htbp]
\centering
\vspace{-0.5cm}
{\footnotesize Without Redshift Tomography:}\\
\vspace{-0.2cm}
\includegraphics[width=16 cm]{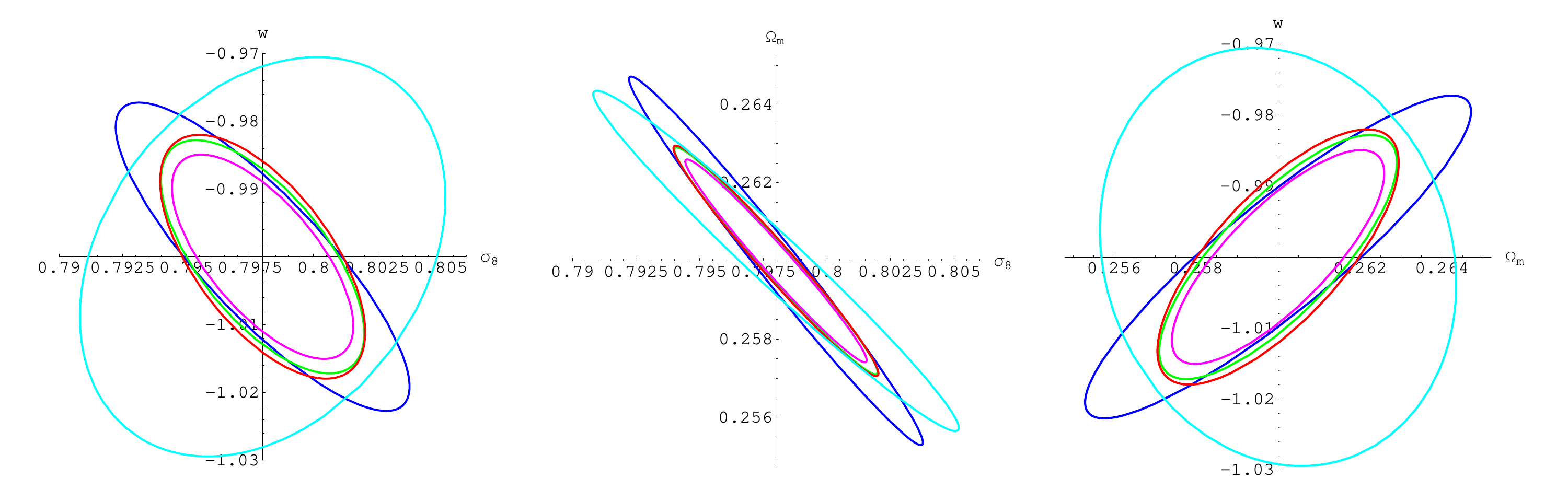}\\
\includegraphics[width=16 cm]{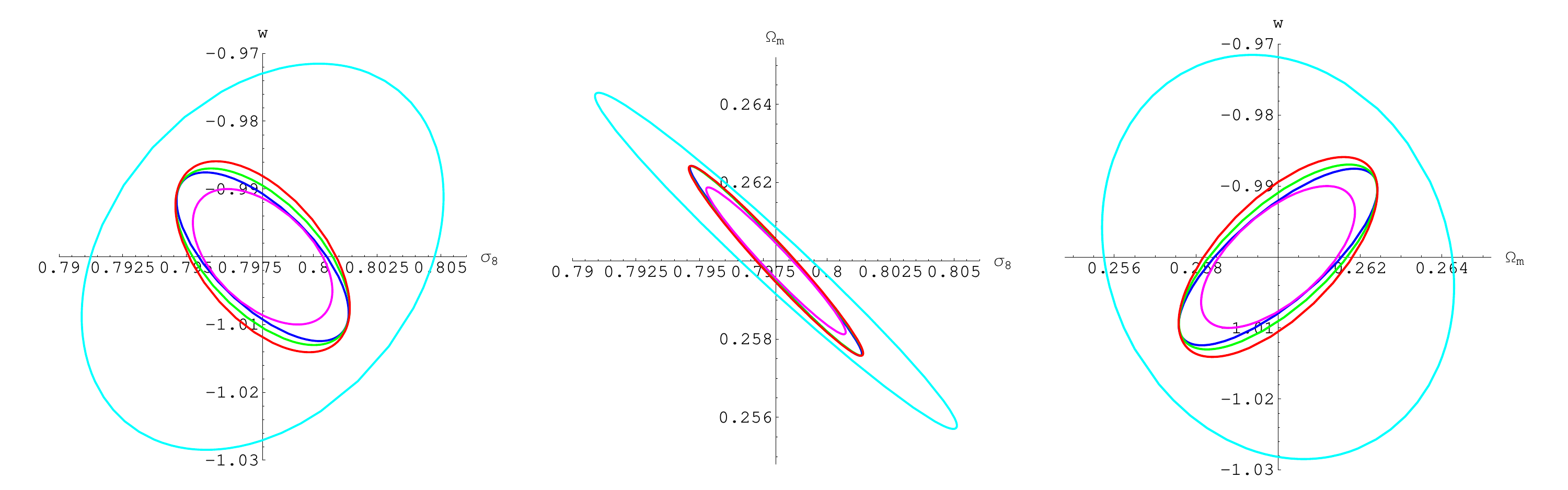}\\
{\footnotesize With Redshift Tomography:}\\
\vspace{-0.2cm}
\includegraphics[width=16 cm]{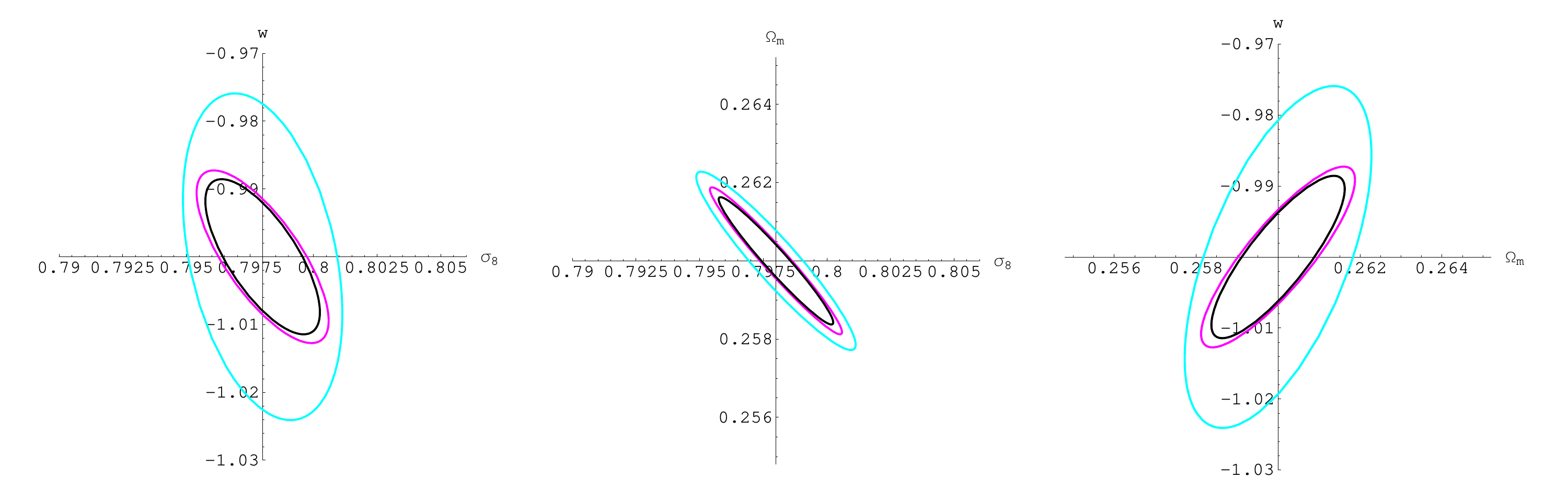}
\includegraphics[width=16 cm]{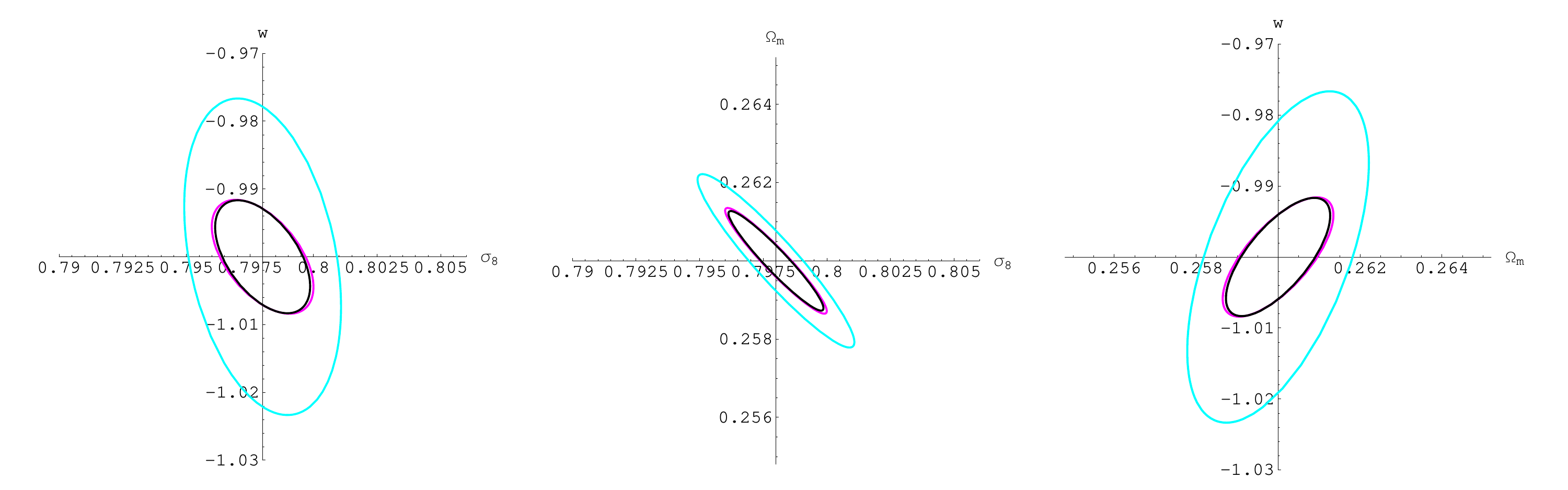}
\hfill
\caption[]{\textit{Comparison of constraints from the Minkowski
    Functionals ($V_i$) and the convergence power spectrum, scaled to
    a full sky survey with a source galaxy density of
    $n_{gal}=15$/arcmin$^2$ per source redshift plane. The colors
    denote: $V_0$ (blue), $V_1$ (green), $V_2$ (red), $V_0$, $V_1$,
    and $V_2$ combined (pink), power spectrum (turquoise), and---only
    in the lower panels---MFs combined with the power spectrum
    (black). The ellipses depict 68.4\% confidence contours,
    marginalized over the third parameter not shown in each panel.
    The top two rows show the constraints from a single redshift
    $z_s=2$, with either a single smoothing scale $\theta_G=1'$ (first
    row) or all five smoothing scales $\theta_G=1', 2', 3', 5',10'$
    combined (second row). The lower set of two rows show the same,
    except using redshift tomography with $z_s=1, 1.5$ and $2$
    combined. Only three smoothing scales $\theta_G=1', 3',10'$ were
    combined in the lowest row to improve numerical stability.  The
    MFs constrain all cosmological parameters more tightly than the
    power spectrum alone, especially the dark energy equation of state
    parameter $w$ (by a factor of $\sim$two). Combining smoothing
    scales tightens the MF constraints much more significantly than
    those from the power spectrum; tomography helps to further tighten
    constraints.  }}
\label{fig:Constraints from Functionals}
\end{figure*}

Figure~\ref{fig:Constraints from Functionals} shows two-dimensional
confidence contours, in each case marginalized over the third
parameter, from the three individual MFs $V_0$ (blue), $V_1$ (green),
and $V_2$ (red), from all three MFs combined (pink), as well as from
the power spectrum alone (turquoise). The lower panels also show the
MFs with the power spectrum combined (black).  The ellipses shown in
this figure enclose 68.4\% of the likelihood, as calculated from the
covariance of the best-fit parameter values with our Monte Carlo
procedure described in Section~\ref{Monte Carlo}.
Table~\ref{tab:Constraints from Functionals} shows the corresponding
68.4\% confidence limits on individual parameters, marginalized over
the other two parameters.

The table and the figure are both scaled from the simulated
12-square-degree maps to the 20,000-square-degree solid angle of a
full-sky survey, such as LSST (i.e.\ by a factor of
$\sqrt{12/20,000}\approx 1/40$). All results shown in this section
were computed using the backward finite difference derivative in
(\ref{Taylor}), which in our cases yielded slightly wider constraints
than the forward derivative. We explore the difference between the two
derivative types explicitly in Sec.~\ref{Accuracy}.

\begin{table}[htbp]
\begin{center}

\begin{tabular}{|c||c|c|c|}
\hline
  & $\Delta\Omega_m$ & $\Delta w$ & $\Delta\sigma_8$ \\
\hline
$z_s=2$; $\theta_G=1'$\\ 
\hline
$V_0$ & 0.00317 & 0.0152 & 0.00393 \\ 
\hline
$V_1$ & 0.00191 & 0.0111 & 0.00263 \\ 
\hline
$V_2$ & 0.00187 & 0.0118 & 0.00262 \\ 
\hline
PS & 0.00297 & 0.0193 & 0.00478 \\ 
\hline
MFs & 0.00175 & 0.00979 & 0.00237 \\
\hline
$z_s=2$; $\theta_G=1',2',3',5',10'$\\ 
\hline
$V_0$ & 0.00153 & 0.00846 & 0.00215 \\ 
\hline
$V_1$ & 0.00163 & 0.0087 & 0.00226 \\ 
\hline
$V_2$ & 0.00158 & 0.00931 & 0.00228 \\ 
\hline
PS & 0.00288 & 0.0189 & 0.00475 \\ 
\hline
MFs & 0.00121 & 0.00668 & 0.00183 \\ 
\hline
$z_s=1,1.5,2$; $\theta_G=1'$\\ 
\hline
$V_0$ & 0.00174 & 0.011 & 0.00204 \\ 
\hline
$V_1$ & 0.00141 & 0.00982 & 0.00188 \\ 
\hline
$V_2$ & 0.00135 & 0.00998 & 0.00183 \\ 
\hline
PS & 0.00156 & 0.0159 & 0.00206 \\ 
\hline
MFs & 0.00122 & 0.00846 & 0.00174 \\ 
\hline
$z_s=1,1.5,2$\\ $\theta_G=1',2',3',5',10'$\\ 
\hline
$V_0$ & 0.000958 & 0.0064 & 0.00143 \\ 
\hline
$V_1$ & 0.000916 & 0.00634 & 0.00131 \\ 
\hline
$V_2$ & 0.00095 & 0.00642 & 0.0014 \\ 
\hline
PS & 0.0015 & 0.0151 & 0.00206 \\ 
\hline
MFs & 0.000912 & 0.00552 & 0.00144 \\ 
\hline
\end{tabular}
\caption[]{\textit{68.4\% confidence limits on cosmological parameters
    from the three Minkowski Functionals and from the power spectrum,
    marginalized over the other two varied parameters---with and
    without combining smoothing scales and with and without using
    tomography, as indicated in the table. Intrinsic ellipticity noise
    from source galaxies with a surface density of
    $n_{gal}=15$/arcmin$^2$ on each redshift plane has been
    included. The numbers have been scaled from our 12-square-degree
    maps to a full-sky LSST-like survey.}}
\label{tab:Constraints from Functionals}
\end{center}
\end{table}

Figure~\ref{fig:Constraints from Functionals} has four rows. In each
row, the 3 panels show the 3 projections of the likelihood ellipsoid
in the $\Omega_m,w,\sigma_8$ space.  The top two rows show the
constraints from a single redshift $z_s=2$, with either a single
smoothing scale $\theta_G=1'$ (first row) or all five smoothing scales
$\theta_G=1', 2', 3', 5',10'$ combined (second row). The lower set of
two rows show the same, except using redshift tomography with $z_s=1,
1.5$ and $2$ combined. We use only three smoothing scales
$\theta_G=1', 3',10'$ for the last row of the figure, because for the
power spectrum it does not make any difference and for the combined
MFs it gives slightly tighter constraints by a few percent for the
numerical reasons described in Sec.~\ref{Maximum Size of the
  Covariance Matrix}.\footnote{The small difference can be seen by
  comparing the last row of Table~\ref{tab:Constraints from
    Functionals} with the middle row of Table~\ref{tab:MF-all}.} The
individual MFs show very similar constraints to each other and keep
the same relative size to the combined-MF contour as in the second
row, so we retain them only in Table \ref{tab:Constraints from
  Functionals} for the redshift tomography case. For the individual
MFs, the constraints are tightest with all smoothing scales combined,
even in the case of redshift tomography. We will comment on the black
ellipse in the third and fourth row of Figure~\ref{fig:Constraints
  from Functionals} in Sec.~\ref{MFs combined with Power Spectrum}.

These are the main results of this paper, and we next turn to
interpreting them further.

\subsection{Information Beyond the Power Spectrum}

A comparison of the pink ellipse (MFs combined) with the turquoise
(power spectrum) in Figure~\ref{fig:Constraints from Functionals}
shows that the MFs constrain all three parameters $\Omega_m$, $w$, and
$\sigma_8$ more tightly than the power spectrum. The tightest
constraints are obtained when all five smoothing scales are combined
and redshift tomography is included (the bottom row in the figure).
The \lq\lq MFs\rq\rq\ and \lq\lq PS\rq\rq\ rows in
Table~\ref{tab:Constraints from Functionals} demonstrate the same
result for the marginalized errors on the individual parameters. While
$\Omega_m$ and $\sigma_8$ are constrained by the MFs only modestly
better, there is a much more significant improvement on the $w$
constraint, by a factor of $\approx$three. Therefore, the MFs appear
particularly useful to improve constrains on dark energy.\footnote{At
least for the simple models studied here with a constant $w$. We plan
to study evolving $w=w(z)$ models in the future.}

We would next like to understand the origin of these constraints
better.  Looking at the leftmost panel of the first row in
Figure~\ref{fig:Constraints from Functionals}, we see that the power
spectrum has a degeneracy in the $(w,\sigma_8)$--direction.  This
degeneracy appears consistent from the changes in the power spectra in panel 2 in
Figure~\ref{fig:power spectrum diff}, which shows that for $\sigma_8$
fixed, the overall normalization of the power spectrum is a bit lower
for the $w=-0.8$ model than for the fiducial cosmology, and therefore
$\sigma_8$ in the $w=-0.8$ model would need to be increased slightly
to make it look more like the fiducial model power
spectrum. This explanation, however, ignores the fact that in Figure~\ref{fig:Constraints from Functionals}
a third parameter is varied, $\Omega_m$. Further, we note that in addition to the overall
  growth, geometrical distance factors contribute importantly to the
  overall $w$-sensitivity of the convergence power spectrum,
  e.g.~\cite{Zhan2009}. 

Interestingly, the MFs show an even stronger degeneracy in the
$(w,\sigma_8)$ plane, but in a nearly orthogonal direction.  All three
MFs place tights constraints in the direction of the degeneracy of the
power spectrum, even without tomography or combining multiple
smoothing scales.  It is instructive to further examine the behavior
of the constraints from $V_0$, with and without combining smoothing
scales.  As mentioned above, in the case of a single smoothing scale,
$V_0$ is equivalent to the fractional area statistic (or histogram),
but when multiple scales are combined, $V_0$ receives, additionally,
spatial information.  We see that combining several smoothing scales
tightens the $V_0$ constraints only modestly in the
$(w,\sigma_8)$--correlation direction, but results in a significant
improvement in the orthogonal, $(w,\sigma_8)$--anticorrelation
direction.  We conclude that in this $(w,\sigma_8)$--anticorrelation
direction, the non-Gaussian information in the maps is coming from
beyond the histogram, whereas in the direction of
$(w,\sigma_8)$--correlation, most of the information is already
contained in the histogram.  {\em This suggests that most of the value
  of the MFs, which helps break degeneracies of $w$ with other
  parameters in the power spectrum, is contained in $V_0$, and the
  single smoothing scale $\sim 1$arcmin is sufficient to get most of
  the benefits.} We elaborate on this finding further in
Section~\ref{Smoothing Scales} below.

\subsection{Redshift Tomography}
\label{Redshift Tomography}

As is well--known, varying either $\Omega_m$ or $w$ changes both the
expansion history of the universe and the redshift dependence of the
amplitude of the power spectrum. In contrast, $\sigma_8$ changes only
the overall normalization of the power spectrum, without modifying its
redshift evolution (at least in the linear regime).  For a single
redshift, the amplitude of the matter fluctuations $\sigma_8$ can be
tuned to compensate for the changes in $\Omega_m$ or $w$, but this
degeneracy is broken when several redshifts are considered
simultaneously.

As Table~\ref{tab:Constraints from Functionals} shows, redshift
tomography is especially useful to break the $(\Omega_m,
\sigma_8)$--degeneracy in the constraints from the power spectrum,
improving constraints on $\Omega_m$ and $\sigma_8$ by factors of 2--3.
The tomographic improvements for $w$, and on all three parameters from
the MFs, are somewhat more modest ($\sim20-30$\%).  Tomography affects
the constraints somewhat more weakly for the MFs, in part because the
$(\Omega_m, \sigma_8)$--degeneracy is weaker to begin with.
Interestingly, Fig.~\ref{fig:Constraints from Functionals} shows
further that in the $(w,\sigma_8)$ plane, tomography tightens the
power spectrum constraints almost entirely in the $\sigma_8$
direction, leaving the constraint on $w$ relatively unimproved.  This
is why, in comparison, the MFs provide a factor of $\sim$three better
marginalized constraint on $w$, even after redshift-tomography is
included.

\subsection{Combining Smoothing Scales}
\label{Smoothing Scales}

Combining smoothing scales turns out to be crucial for MFs -- even
more important than tomography. This was already anticipated in
\cite{whm09}, and is here clearly evidenced by the tighter constraints
in the 2nd and 4th row in Figure~\ref{fig:Constraints from
  Functionals}, where multiple smoothing scales are combined. This is
in sharp contrast with the power spectrum, where combining smoothing
scales results in negligible improvements (as it should, at least in
the linear regime).  The improvements are most pronounced in the
constraints from $V_0$, especially for a single redshift, as can be
seen by comparing the upper left panel with the left panel in the
second row. We thus further examine this case, as an example of how
MFs derive information from smoothing scale combinations.

Consider the $(w,\sigma_8)$--anticorrelation direction in the top left
panel of Figure~\ref{fig:Constraints from Functionals}. As we have
noted above, the power spectrum constraints are strongest in this
direction. The power spectrum measures spatial correlations, and $V_0$
is completely blind to spatial arrangements of pixel values, when only
a single smoothing scale is used. It is therefore unsurprising to see
that $V_0$ constraints in this direction are particularly weak.  The
other two MFs -- both sensitive to aspects of shape and topology --
incorporate some spatial information, and fare much better. However,
when we combine multiple smoothing scales, the $V_0$ constraint
tightens to the same level as the other MFs (see left panel in the
second row). This makes intuitive sense, since combining smoothing
scales adds spatial distribution information to $V_0$: Whether peaks
on smaller scales are clustered together determines if they survive
large-scale smoothing or get washed out by it.

The above interpretation can be further illuminated by considering the
theoretical predictions for the MFs for GRFs. We see from
Eqs.~(\ref{sigma0})--(\ref{sigma1}) that $V_0$ depends only on
$\sigma_0$, defined in Eq.~(\ref{sigma0}). This is a simple integral
of the power spectrum, and is insufficient to capture all the
cosmological information in the $(w, \sigma_8)$--anticorrelation
direction. In comparison, $V_1$ and $V_2$ depend, additionally, on
$\sigma_1$, the first moment of the power spectrum defined in
Eq.~(\ref{sigma1}), which weighs high $\ell$ more than $\sigma_0$. In
combination, these two integrals provide additional cosmological
information from the shape of the power spectrum.  When combining
different smoothing scales, the power spectrum is truncated at
different values of $\ell$.  As a result, the combination of
$\sigma_0$'s from different smoothing scales captures information from
the $\ell$--dependence of the power spectrum, providing $V_0$ with
information that is similar to $V_1$ and $V_2$.

This raises the question: How many smoothing scales does one need?
Combining too many smoothing scales increases the size of the
covariance matrix and can render the results unreliable, as discussed
above. Table~\ref{tab:Effect of smoothings} shows the marginalized
constraints from different individual smoothing scales and from their
combinations for all three MFs combined.

Unsurprisingly, the smallest smoothing scale, $\theta_G=1$~arcmin,
provides the tightest constraints, and the constraints get
progressively weaker for larger smoothing scales, up to a factor of
$\sim2$ for $\theta_G=10$~arcmin.  Nevertheless, there is
complementary information in these larger scales.  Interestingly, the
table shows that one needs to combine at least three scales to get
most of the improvement shown by the combination of all five scales we
studied. (Since we have not probed more than five smoothing scales, we
cannot say whether adding even more scales may further improve the
result.)

\begin{table}[htbp]
\begin{center}

\begin{tabular}{|c||c|c|c|}

\hline
 & $\Delta\Omega_m$ & $\Delta w$ & $\Delta\sigma_8$ \\
\hline
$\theta_G=1'$ & 0.00175 & 0.00979 & 0.00237 \\ 
\hline
$\theta_G=2'$ & 0.00199 & 0.0106 & 0.00268 \\ 
\hline
$\theta_G=3'$ & 0.00201 & 0.0108 & 0.00283 \\ 
\hline
$\theta_G=5'$ & 0.00225 & 0.0127 & 0.00316 \\ 
\hline
$\theta_G=10'$ & 0.00322 & 0.0152 & 0.00468 \\ 
\hline
$\theta_G=1',3'$ & 0.00155 & 0.00849 & 0.00211 \\ 
\hline
$\theta_G=1',5'$ & 0.00149 & 0.00816 & 0.00212 \\ 
\hline
$\theta_G=1',2',3'$ & 0.00142 & 0.00757 & 0.00197 \\ 
\hline
$\theta_G=1',3',10'$ & 0.00139 & 0.00735 & 0.00196 \\ 
\hline
$\theta_G=1',2',3',5',10'$ & 0.00121 & 0.00668 & 0.00183 \\ 
\hline

\end{tabular}
\caption[]{\textit{Marginalized constraints on cosmological parameters
    from all three MFs combined, for different individual smoothing
    scales and from various smoothing--scale combinations. See the
    caption of Table~\ref{tab:Constraints from Functionals} for a
    further explanation of the numbers in the table. A single source
    redshift $z_s=2$ was assumed. The table shows that combining
    smoothing scales is advantageous for MFs and does not reach a
    minimum for at least five scales.}}
\label{tab:Effect of smoothings}
\end{center}
\end{table}

\subsection{MFs combined with Power Spectrum}
\label{MFs combined with Power Spectrum}

We have so far compared the constraints from MFs with those from the
power spectrum.  Since the constraints from the MFs is stronger, it is
clear that the MFs contain information beyond the power spectrum.
However, another interesting question is: To what extent are the
constraints from the MFs and the power spectrum independent? Could
constraints tighten further when they are combined?

To answer this question, ideally we would like to combine the MFs with
the power spectrum, using all five smoothing scales, as well as
redshift tomography. Unfortunately, as discussed above, the covariance
matrix in this case exceeds the maximum reliable size allowed by our
simulated datasets.

However, we have found that combining only three smoothing scales, the
results remain stable (in the sense discussed in \S\ref{Maximum Size
  of the Covariance Matrix} above) even when tomography with all three
redshifts $z_s=1, 1.5, 2$ is used. Table~\ref{tab:MF-all} and the last
row of Figure~\ref{fig:Constraints from Functionals} show the results
in this case. The figure shows the projected error ellipses from the
power spectrum only (turquoise), from all three MFs combined (pink),
and from the combination of all three MFs and the power spectrum
(black); the table shows the corresponding marginalized constraints on
the individual parameters. Clearly, adding the power spectrum does not
yield any improvement on the constraints already available from the
MFs.

Since we are operating at the quality limit of our dataset, we
evaluated several other combinations with smaller covariance
matrices. The third row of Figure~\ref{fig:Constraints from
  Functionals} shows redshift tomography with only one smoothing
scale. At first glance, here adding the power spectrum to the MFs
seems to show a small advantage: The contours get 11\% tighter for
$\Omega_m$ and 14\% tighter for $w$ and $\sigma_8$. However, this is
because only a single smoothing scale is used, which does not extract
the maximum amount of information from the MFs. If one studies only
one redshift and combines all five smoothing scales, again nothing is
gained from adding the power spectrum to the MFs.

This paints a complete picture. \emph{We conclude that as long as
  several smoothing scales are combined, the MFs already extract all
  the information which is in the power spectrum. }

\begin{table}[htbp]
\begin{center}

\begin{tabular}{|c||c|c|c|}
\hline & $\Delta\Omega_m$ & $\Delta w$ & $\Delta\sigma_8$ \\ 
\hline 
PS & 0.00147 & 0.0151 & 0.00209 \\ 
\hline
MFs & 0.000858 & 0.00549 & 0.00126 \\ 
\hline
MFs+PS & 0.000815 & 0.00553 & 0.0012 \\
 \hline
\end{tabular}
\ \\
\caption[]{\textit{Marginalized constraints on cosmological parameters
    from the power spectrum, the three Minkowski functionals (MFs)
    combined, and from the power spectrum together with the
    MFs. Redshift tomography with source planes at $z_s=1, 1.5, 2$ and
    a combination of three smoothing scales $\theta_G=1', 3', 10'$
    were used (using all five smoothing scales with tomography would
    have caused the constraints to widen by a few percent for
    numerical reasons, see text). Intrinsic ellipticity noise from a
    source galaxy surface density of $n_{gal}=15$/arcmin$^2$ per
    redshift plane has been included. The table shows that the MFs
    already include all of the information that is in the power
    spectrum: Adding the power spectrum does not improve the
    constraints further. These are our tightest constraints.}}
\label{tab:MF-all}
\end{center}
\end{table}

\subsection{Noiseless Minkowski Functionals}

To illustrate the effect of ellipticity noise on our constraints, we
repeated our analysis on noiseless maps, for all three redshifts
$z_s=1, 1.5, 2$ and three smoothing scales $\theta_G=1', 3', 10'$
combined, which constitutes our best numerically still stable case.
For reference, we note that the r.m.s.\ fluctuations in $\kappa$
caused by large-scale structures at $z_s=2$ are
$\sigma_{\kappa}=0.022$, very close to the r.m.s.\ noise $\sigma_{\rm
  noise}=0.023$ added to the maps~\cite{Yang:2011zz}.  We thus expect
noise to degrade the constraints by of order a factor of $\sim$two.

\begin{figure*}[htbp]
\centering 
\includegraphics[width=16 cm]{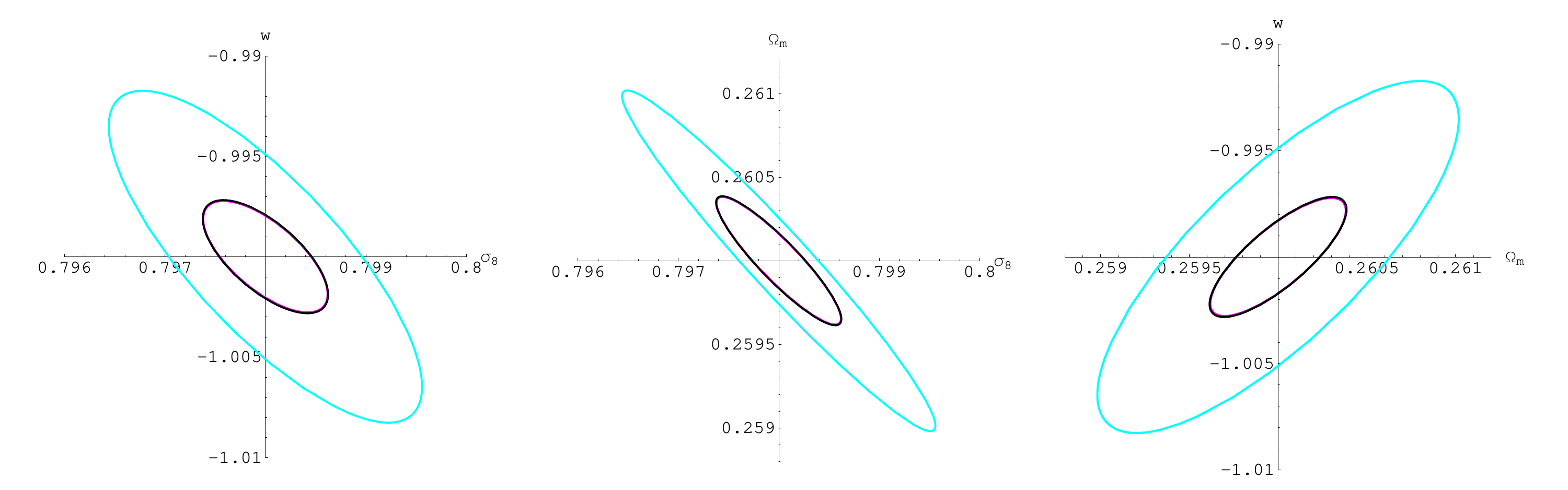} 
\hfill
\caption[]{\textit{Comparison of 68.4\% confidence ellipses from maps
    without intrinsic ellipticity noise. This figure is the noiseless
    equivalent of the last row of Figure~\ref{fig:Constraints from
      Functionals}.  The colors are the same as in that figure.  The
    figure shows that noise hurts the MFs more than the power
    spectrum: the MFs (pink) have a larger advantage---a factor of
    almost $\approx 3$--- over the power spectrum (turquoise) than in
    the noisy maps. As in the case with noise, adding the power
    spectrum to the MF constraints (black) yields no further
    improvement.}}
\label{fig:Noiseless Constraints}
\end{figure*}

The results are displayed in Table~\ref{tab:Noiseless Constraints} and
Figure~\ref{fig:Noiseless Constraints}.  As before, the figure shows
the constraints from the power spectrum alone (turquoise), the three
MFs combined (pink) and the combination of the MFs and the power
spectrum (black).

The table and the figures demonstrate that the constraints are tighter
than in the noisy case. We also find that noise hurts the power
spectrum somewhat more than the MFs: The degradation due to noise for
the marginalized constraint on $\Omega_m, w$, and $\sigma_8$ is a
factor of 2.1, 2.8, and 1.8 for the power spectrum (close to the
expected factor of $\sim$two), whereas the corresponding degradations
for the MFs are factors of 3.4, 3.0, and 3.2.  As a result, in the
noiseless case, the MFs outperform the power spectrum by a factor of
$\sim 3$, larger than in the noisy case.

\begin{table}[htbp]
\begin{center}

\begin{tabular}{|c||c|c|c|}

\hline
 & $\Delta\Omega_m$ & $\Delta w$ & $\Delta\sigma_8$ \\
\hline
PS & 0.00067 & 0.00534 & 0.00105 \\ 
\hline
MFs & 0.00025 & 0.00184 & 0.000408 \\ 
\hline
MFs+PS & 0.00025 & 0.00183 & 0.000403 \\ 
\hline
\end{tabular}
\caption[]{\textit{Comparison of marginalized constraints from the
    MFs, the power spectrum, and from their combination, but without
    adding any intrinsic ellipticity noise.  This table is the
    noiseless equivalent of Table~\ref{tab:MF-all}.
  }}\label{tab:Noiseless Constraints}
\end{center}
\end{table}

\subsection{Comparison to Previous Results}
\label{Comparison to Results}

\begin{figure}[htbp]
\centering
\includegraphics[width=8 cm]{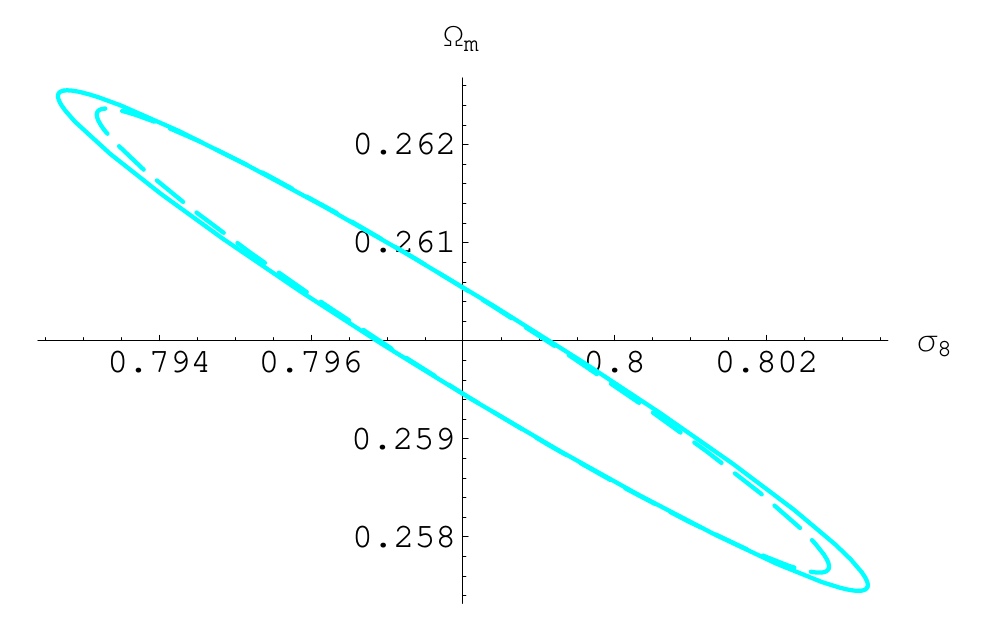}
\hfill
\caption[]{\textit{Error ellipse from the convergence power spectrum
    alone in the ($\Omega_m,\sigma_8$) plane, with $w$ held
    fixed. This is to be compared to Figure~5 in
    ref.~\cite{Berge:2009xj}, who use $n_{gal}=40$/arcmin$^2$ and a
    source galaxy redshift distribution around $z_s=1$ for a different
    cosmology $(\Omega_m,\sigma_8)=(0.3, 0.9)$. To resemble the
    parameters of \cite{Berge:2009xj} as closely as possible, we
    adopted $n_{gal}=40$/arcmin$^2$ and $z_s=1$, but use our fiducial
    cosmology.  The dashed [solid] contour uses forward [backward]
    finite difference derivatives.  The agreement of our ellipse with
    Figure 5 in \cite{Berge:2009xj} is very good, validating our
    pipeline and results.}}
\label{fig:Ellipse like Berge}
\end{figure}

To validate our simulation pipeline, as well as our results, we next
compare the results of our Fisher matrix analysis for the power
spectrum to a recent theoretical study \cite{Berge:2009xj}.  While
there are many other previous estimates for cosmological constraints
from the WL power spectrum (e.g.\ \cite{S&K04}, and for a very recent
study, examining constraints from power spectrum of the logarithm of
the convergence field, see \cite{Seo2011}) the specifications in the
study by \cite{Berge:2009xj} are closest to the present
work.  In particular, those authors present a two--dimensional Fisher
ellipse, in Figure~5 of their paper, in which they vary only the two
parameters $\Omega_m$ and $\sigma_8$ . Their results were computed for
a source galaxy redshift distribution around $z_s=1$ and surface
density of $n_{gal}=40$/arcmin$^2$.  We have therefore rerun our
analysis for the same source galaxy surface density, and plot our
power spectrum constraints in Figure~\ref{fig:Ellipse like Berge} for
$z_s=1$.  We have also kept $w$ fixed at its fiducial value of $w=-1$.
The solid (dashed) ellipse uses backward
(forward) finite differences. The agreement between both the
orientation and overall size of our error ellipse and that shown in
Figure 5 of \cite{Berge:2009xj} is very good, especially considering
that theirs is for a different cosmology with
$(\Omega_m,\sigma_8)=(0.3, 0.9)$. Ours is a factor of 1.25 larger in
$\Omega_m$ and a factor of 1.4 in $\sigma_8$.

\subsection{Accuracy}
\label{Accuracy}

Here we discuss the variability of the error contours under different
sets of maps for the fiducial model, to get an estimate of how much
our results depend on the particular set of simulations and
realizations of individual lensing maps. We then enumerate other
possible caveats and systematic errors that we did not explicitly take
into account.

\subsubsection{Uncertainties Explicitly Evaluated}

We have two strictly independent sets of 1,000 maps in the fiducial
model, obtained using different sets of N-body runs.  We thus have a
choice of which of these sets to use for (i) computing the Simulation
Mean (eq.~\ref{SM}), and (ii) computing the covariance matrix
(eq.~\ref{covariance matrix}), and (iii) minimizing $\chi^2$ and Monte
Carlo.  In all of our results so far, we used the auxiliary set for
the Simulation Mean and the covariance matrix, and the fiducial set
for Monte Carlo.  We repeated our calculations by swapping the map
sets used for the covariance matrix and for Monte Carlo, and confirmed
that this has a negligible effect on our results.  Additionally, we
have found that for the MFs the results are almost independent of the
set used for the Simulation Mean. Although we found a dependence on
which set is used for the Simulation Mean for the power spectrum, this
is modest: The change is smaller by a factor of 3--4 than the dominant
source of variability described in the next paragraph and expresses
itself mostly as a slight rotation of the ellipse rather than a change
in its size.

The largest numerical variability, however, comes from the difference
between the forward vs.\ backward finite difference used for the
parameter dependence of the descriptors in the Taylor expansion
(eq.~\ref{Taylor}), which can be taken as an indication of the
robustness of our constraints. Evidently, our choice of parameter
spacing between the simulations was somewhat outside the linear
regime, such that these two differences yield different derivatives.
This is the dominant source of uncertainty.  We decided against a
higher-order interpolation scheme to improve on our analysis because
these variations are modest in most cases, and furthermore, they
affect the power spectrum and the MF results in similar
ways. Therefore, we have confidence that the relative strength of our
constraints between the power spectrum and the MFs holds accurately.

\begin{figure*}[htbp]
\centering
{\footnotesize Without Redshift Tomography:}\\
\vspace{-0.2cm}
\includegraphics[width=16 cm]{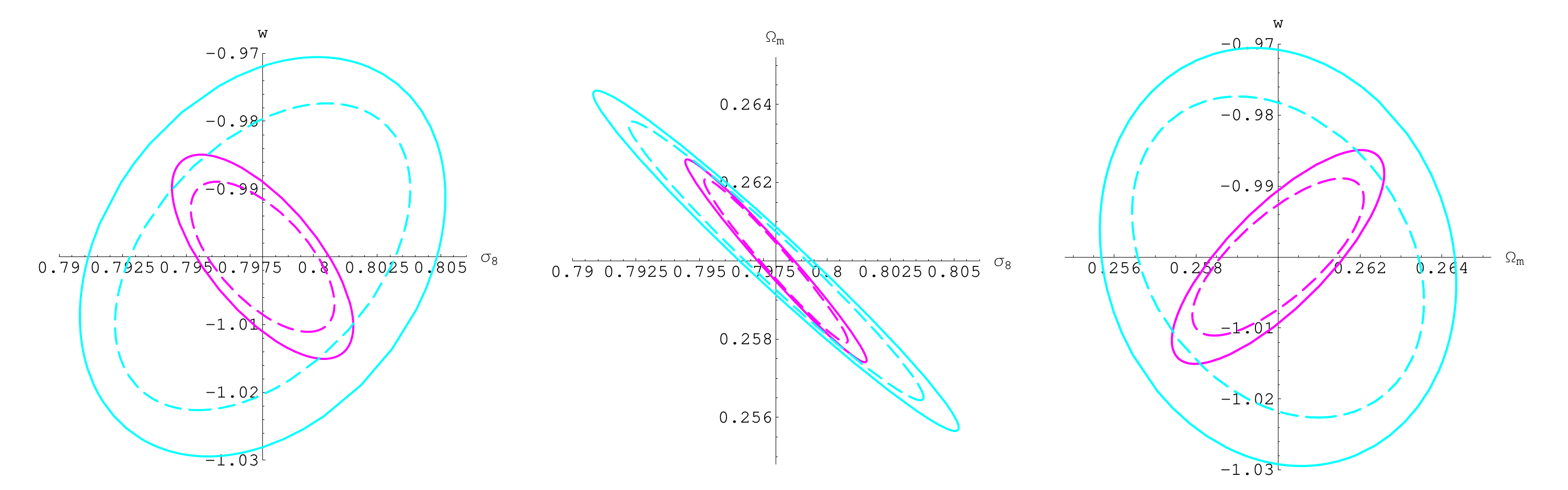}
\hfill
\caption[]{\textit{Illustration of the largest source of variability
    in our numerical results. In addition to showing the constraints
    from using the backward finite difference derivatives (solid) in
    the Taylor expansion Eq.~(\ref{Taylor}) as in the other figures,
    here we also show the ellipses computed from the forward
    derivatives (dashed). Otherwise this figure is the same as the
    first row of Figure~\ref{fig:Constraints from Functionals} for
    $z_s=2$ and $\theta_G=1$ arcmin, but showing only the constraints
    from the MFs combined (pink) and the convergence power spectrum
    (turquoise). The differences between the ellipses from different
    derivatives are small enough not to impact our conclusions, and
    tend to get even smaller as more redshifts and smoothing scales
    are combined. Furthermore, the relative differences between the
    MFs and power spectrum are well preserved in both cases
    individually.}}
\label{fig:Derivatives}
\end{figure*}

We illustrate the effect of using the two different derivatives on the
contours from MFs and the power spectrum in
Figure~\ref{fig:Derivatives} for one redshift $z_s=2$ and one
smoothing scale $\theta_G=1$ arcmin. This figure is the equivalent of
the first row in Figure~\ref{fig:Constraints from Functionals}, but
with both derivative cases shown. The solid and dashed curves in the
figure indicate the constraints calculated from backward and forward
finite differences in eq.~(\ref{Taylor}) between our simulations,
respectively. The differences between the ellipses in the figure are
of a representative size: The difference between the contours from the
two derivatives tends to get smaller as more redshifts and smoothing
scales are combined. In all our other figures and tables we choose to
show only the more conservative of the two: results using the backward
derivative. The difference in the error contours is still small enough
so that it does not affect our overall conclusions. The errors due to
the choices of map sets are small in comparison and are not shown.

\subsubsection{Other Sources of Error}\label{Caveats}

Our study was aimed at a theoretical exploration of the cosmological
information content and utility of the MFs. As such, we made a list of
idealized assumptions -- we here present a partial list of neglected
sources of systematic errors, which will have to be quantified in
future work.

We used convergence in our maps, rather than the reduced shear that
will be available directly from the observations.  One can, in
principle, convert one quantity to the other; in practice, on small
scales, where the MFs derive most of their constraining power, the
errors introduced by this conversion will not be negligible, and must
be taken into account when extracting cosmological parameters from
actual data (e.g. \cite{Dodelson:2005ir}).

We assumed the noise from the intrinsic ellipticities of the source
galaxies is given by independent Gaussian distributions, with a single
width determined by the assumed average source galaxy density. This
neglects effects such as shot noise from random galaxy positions,
correlations due to intrinsic alignments of galaxies, and
magnification bias.  In the future, ellipticity noise should be
incorporated more accurately, using a mock galaxy catalog with random
galaxy positions and intrinsic ellipticity components drawn from more
realistic (non-Gaussian) probability distributions, which would then
have to be added to the two independent components of the shear signal
separately.  Additionally, a realistic redshift-distribution of source
galaxies should be employed, and photometric redshift errors folded in
the analysis, rather than confining all galaxies to fixed redshift
planes.

Furthermore, there are systematic effects from the atmosphere and the
instrument of a real telescope on one hand, as well as holes in the
maps due to foreground stars and other impurities, which we have not
taken into account.

The small simulation box size causes a fall--off in the power spectrum
on large scales, and we were unable to study correlations between
fields larger than 12 square degrees, assuming them to be independent
in our idealized full-sky scaling.

Finally, the variance in our MF bins may be somewhat underestimated,
because we use only 45 independent N-body simulations to create our
pseudo-independent 1000 convergence maps for the fiducial cosmology,
and only 5 N-body simulations for the maps of the other
cosmologies. 
We plan to create a larger suite of N-body simulations,
and study the importance of the number of independent runs required
for weak lensing map generation, in a future paper.  On the positive
side, a larger suite of runs would allow us to study the dependence of
the (co)variances $C_{ij}$ themselves on the background cosmology, and
to assess whether these add significant extra information.

%%%%%%%%%%%%%%%%%%%%%%%%%%%%%%%%%%%%%%%%%%%%%%%%%%%%%%%%%%%%%%%%%%%%%%%%%%%%%%%%%%%%%%%%%%
\section{Conclusions} \label{Conclusions}
%%%%%%%%%%%%%%%%%%%%%%%%%%%%%%%%%%%%%%%%%%%%%%%%%%%%%%%%%%%%%%%%%%%%%%%%%%%%%%%%%%%%%%%%%%

We have studied the cosmological information content of Minkowski
functionals (MF) derived from mock weak lensing maps from an extensive
suite of ray-tracing N-body simulations.  While there have been a few
smaller precursor works, applying MFs to weak lensing maps, this is
the first large systematic study to our knowledge.  Our N-body
simulations cover seven different cosmological models, bracketing the
parameters $\Omega_m$, $\sigma_8$, and $w$ around a fiducial
$\Lambda$CDM cosmology.  We created convergence maps with ray-tracing,
added intrinsic ellipticity noise from the source galaxies, and
obtained joint confidence limits on parameters with a Monte Carlo
procedure.

Our main result is that there is a substantial amount of information
from non-Gaussian features in WL maps, which, as we have explicitly
verified, is coming from beyond the power spectrum.  In particular,
the constraints on the dark energy equation of state parameter, $w$,
marginalized over $\Omega_m$ and $\sigma_8$, is nearly a factor of three
tighter than from the power spectrum alone. From combining the MFs
with the power spectrum, we also demonstrated that the MFs contain all
of the information available from the power spectrum.

The non-Gaussian information extracted by the MFs resides in part in
the one-point function of the convergence, which places constraints
primarily in the $(w, \sigma_8)$--correlation direction, and helps
break the strongest degeneracy between these parameters present in the
power spectrum.  When multiple smoothing scales are combined, the MFs
derive further information from the morphology and topology of the
iso-convergence contours, and place tight constraints in the
orthogonal $(w, \sigma_8)$--anticorrelation direction.  We further
find that redshift tomography is important to break the degeneracy
between $\Omega_m$ and $\sigma_8$.  However, the marginalized
constraint on $w$ from the power spectrum alone remains much less
affected by tomography.

It would be interesting to study how the constraints from Minkowski
functionals complement those from the power spectrum under more
realistic survey conditions---with improved theoretical modeling, and
including instrument and atmospheric systematics.  The MFs will be
available automatically in future WL survey data, and our results
suggest that they will improve cosmological constraints. Their
treatment here is an important step towards realizing the full
potential of weak lensing maps, including the information from
nonlinear structures.

%%%%%%%%%%%%%%%%%%%%%%%%%%%%%%%%%%%%%%%%%%%%%%%%%%%%%%%%%%%%%%%%%%%%%%%%%%%%%%%%%%%%%%%%%%
\section{Acknowledgments}
%%%%%%%%%%%%%%%%%%%%%%%%%%%%%%%%%%%%%%%%%%%%%%%%%%%%%%%%%%%%%%%%%%%%%%%%%%%%%%%%%%%%%%%%%%

We thank Volker Springel for useful discussions and for providing the
N-body initial conditions generator N-GenIC, Lam Hui for providing us
with a $w$-capable linear growth factor code, and Xiuyuan Yang and Dennis Simon
for helpful discussions. We also thank Leonard Slatest
and Efstratios Efstathiadis for help with the IBM Blue Gene at BNL.

JMK and KMH acknowledge support from JPL through subcontract 1363745.
This research utilized resources at the New York Center for
Computational Sciences at Stony Brook University/Brookhaven National
Laboratory which is supported by the U.S. Department of Energy under
Contract No. DE-AC02-98CH10886 and by the State of New York. Almost
all the calculations were performed on the IBM Blue Gene/L and /P New
York Blue. The matter power spectra for initial conditions and the
$\chi^2$--minimization procedure were calculated on the LSST/Astro
Linux cluster at BNL.

\vfill

%%%%%%%%%%%%%%%%%%%%%%%%%%%%%%%%%%%%%%%%%%%%%%%%%%%%%%%%%%%%%%%%%%%%%%%%%%%%%%%%%%%%%%%%%%

\end{document}